%% file: natren.tex
\input{natren.def}
\documentstyle[12pt,epsf]{article}\title{Natural renormalization}\author{Oliver Schnetz\thanks
{Institut f{\"u}r theoretische Physik III, Staudtstra{\ss}e 7, 91058 
Erlangen, Germany,\newline 
e-mail: schnetz@pest.physik.uni-erlangen.de\newline 
Supported in parts by the DFG Graduiertenkolleg 'Starke Wechselwirkung' 
and the BMBF.\newline 
FAU-TP3-96/1}}\date{June 18 1996}
\begin{document} 
\maketitle 
\begin{abstract} 
A careful analysis of differential renormalization shows that a distinguished 
choice of renormalization constants allows for a mathematically 
more fundamental interpretation of the scheme. With this set of 
a priori fixed integration constants differential renormalization 
is most closely related to the theory of generalized functions. 
The special properties of this scheme are illustrated by application 
to the toy example of a free massive bosonic theory. Then we apply 
the scheme to the $\varphi ^{4}$-theory. The two-point function 
is calculated up to five loops. The renormalization group is analyzed, 
the beta-function and the anomalous dimension are calculated up 
to fourth and fifth order, respectively.
\end{abstract} 
\tableofcontents 
\section{Introduction} 
With the proof of renormalizability of non-Abelian gauge theories 
in the early 1970's the problem of giving a perturbative definition 
of a renormalizable quantum field theory was solved (eg.\ \cite{Collins}). 
However explicit calculations in the commonly used dimensional regularization 
are often tedious. This kept the interest in alternative prescriptions 
alive.

Quite recently differential renormalization \cite{Johnson,Freedman,Latorre} 
has been proposed. For practical calculations this renormalization 
scheme provides two major advantages. Firstly, it allows to regularize 
and renormalize in one step. No explicit regulators or counterterms 
are needed. Secondly, it is possible to keep the spacetime dimension 
fixed. This is particularly useful for dimension-specific theories 
like the chiral electroweak sector of the standard model.

We start by analyzing differential renormalization from a purely 
mathematical point of view. Differential renormalization is usually 
formulated in four-dimensional coordinate space by writing divergent 
amplitudes as Laplacians (we restrict ourselves to Euclidean signature, 
$ x^{4}=(x^{2})^{2}$, $ \Box =\sum _{i}\partial _{i}^{2}$) of less 
divergent expressions. For example, 
\begin{equation}
\label{1}\frac{1}{x^{4}{}} =-\frac{1}{4}\Box \frac{1}{x^{2}{}} \ln
\left( \frac{x^{2}}{\Lambda _{0}^{2}}\right) 
\hspace{.6ex},\hspace{2ex}\hspace*{1cm}\frac{1}{x^{6}{}} =-
\frac{1}{32}\Box \Box \frac{1}{x^{2}{}} \ln\left( \frac{x^{2}}{
\Lambda _{1}^{2}}\right) \hspace*{2ex}\hbox{ for }x\neq 0
\hspace{.6ex}.
\end{equation} 
$\Lambda _{0}$ and $\Lambda _{1}$ are arbitrary integration constants 
which are kept for dimensional reasons.

Initially ill-defined integrals are now regularized by the convention 
that the Laplacian should act on the left and the surface term is 
ignored. According to this rule the singular Fourier transforms 
of $ x^{-4}$ and $ x^{-6}$ can be derived from the well-defined 
Fourier transform of $ x^{-2}\ln(x^{2}{} /\Lambda ^{2})$ (calculated 
below, Eq.\ (\ref{I})), 
\begin{eqnarray}
\label{2}\int \limits _{{\rm diff.ren.\hspace{.38ex}}}\frac{d^{4
}x}{4\pi ^{2}{}} \frac{e^{ix\cdot p}}{x^{4}{}} &\equiv &-\frac{
1}{4}\int \frac{d^{4}x}{4\pi ^{2}{}} \left( \Box e^{ix\cdot p}
\right) \frac{1}{x^{2}{}} \ln\left( \frac{x^{2}}{\Lambda _{0}^{
2}}\right) \hspace*{1ex}=\hspace*{1ex}-\frac{1}{4}\ln\left( 
\frac{p^{2}}{\bar{\Lambda }_{0}^{2}}\right) \hspace{.6ex},\\ 
\label{2a}\int \limits _{{\rm diff.ren.\hspace{.38ex}}}\frac{d^{
4}x}{4\pi ^{2}{}} \frac{e^{ix\cdot p}}{x^{6}{}} &\equiv &-
\frac{1}{32}\int \frac{d^{4}x}{4\pi ^{2}{}} \left( \Box \Box e^
{ix\cdot p}\right) \frac{1}{x^{2}{}} \ln\left( \frac{x^{2}}{
\Lambda _{1}^{2}}\right) \hspace*{1ex}=\hspace*{1ex}\frac{1}{32
}p^{2}\ln\left( \frac{p^{2}}{\bar{\Lambda }_{1}^{2}}\right) 
\hspace{.6ex},
\end{eqnarray} 
where $ \bar{\Lambda }_{0(1)}=2/(e^C\Lambda _{0(1)})$ and $C$=0.5772156{\dots} 
is the Euler constant.

The central point in this paper is to exhibit the meaning of the 
above prescriptions for one-dimensional integrals. To this end we 
perform the convergent angular integrals in (\ref{2}), (\ref{2a}) 
which leaves us with a radial integral $ \int _{0}^{\infty }dr$ 
that diverges at $ r=0$. We finally split this integral into a convergent 
part $ \int _{1}^{\infty }$ which can be evaluated and a singular 
part $ \int _{0}^{1}$ which is kept. These entirely well-defined 
manipulations lead to 
\begin{eqnarray}
\int \frac{d^{4}x}{4\pi ^{2}{}} \frac{e^{ix\cdot p}}{x^{4}{}} &=
&-\frac{1}{2}\left( \ln\left( \frac{e^C|p|}{2}\right) -\int _{0
}^{1}\frac{dr}{r} -\frac{1}{2}\right) \hspace{.6ex},\\ 
\label{888a}\int \frac{d^{4}x}{4\pi ^{2}{}} \frac{e^{ix\cdot p}
}{x^{6}{}} &=&\frac{1}{16}p^{2}\left( \ln\left( \frac{e^C|p|}{2
}\right) -\int _{0}^{1}\frac{dr}{r} -\frac{5}{4}\right) +\frac{
1}{2}\left( \int _{0}^{1}\frac{dr}{r^{3}{}} +\frac{1}{2}
\right) \hspace{.6ex}.
\end{eqnarray} 
Now we compare this result with Eqs.\ (\ref{2}) and (\ref{2a}) derived 
by differential renormalization. First, notice that the second term 
on the right hand side of Eq.\ (\ref{888a}) has no $ p^{2}$-dependence 
at all. To make the right hand side proportional to $ p^{2}$ we 
define 
\begin{equation}
\label{A}\int _{0}^{1}\frac{dr}{r^{3}{}} \equiv -\frac{1}{2}
\hspace{.6ex}.
\end{equation} 
This leads us finally to the equations 
\begin{equation}
\label{B}\int _{0}^{1}\frac{dr}{r} =-\frac{1}{2}-\ln\Lambda _{0}
=-\frac{5}{4}-\ln\Lambda _{1}\hspace{.6ex}.
\end{equation} 
Eqs.\ (\ref{A}) and (\ref{B}) can be seen as one-dimensional definitions 
of differential renormalization. However Eq.\ (\ref{B}) naturally 
relates the renormalization constants via 
\begin{equation}
\label{C}\ln\left( \frac{\Lambda _{1}}{\Lambda _{0}}\right) =-
\frac{3}{4}\hspace{.6ex}.
\end{equation} 
For a ratio $ \Lambda _{1}/\Lambda _{0}$ different from exp(-$\frac{3}{4}$) 
the one-dimensional interpretation of differential renormalization 
is not possible.

We will see in the next section that all ratios of renormalization 
constants are fixed by consistency conditions. Differential renormalization 
with these a priori fixed ratios will be called 'natural renormalization'.

It was shown \cite{Latorre} that differential renormalization provides 
a self-consistent definition of renormalizable field theories, without 
referring to the ratio $ \Lambda _{1}/\Lambda _{0}$ as given in 
Eq.\ (\ref{C}). Moreover in some cases it is convenient to adjust 
the ratios of renormalization constants according to physical requirements 
\cite{Smirnov2}. In particular for gauge theories it is useful to 
fix some ratios by Ward identities \cite{Johnson,Haagensen,Freedman2}. 
However, depending on the gauge, some of these ratios may differ 
from the prescriptions we give. The treatment of gauge theories 
in natural renormalization is still under investigation, first attempts 
have been successful \cite{Schnetz}.

\vspace{1ex}
\noindent{}The main advantage of allowing for the above one-dimensional 
reduction and demanding Eqs.\ (\ref{A}), (\ref{B}), (\ref{C}) is 
that differential renormalization can be understood on a much more 
general footing. We will see in Sec.\ \ref{genfunct} that Eqs.\ 
(\ref{A}) and (\ref{B}) are almost standard in the theory of generalized 
functions. Thus it becomes possible to replace the recipes of differential 
renormalization by mathematically more fundamental definitions.

In contrast to differential renormalization, natural renormalization 
is neither connected to coordinate nor to momentum space. One has 
the freedom to choose the most convenient representation for the 
respective problem.

The first example where natural renormalization becomes advantageous 
is the toy theory of free massive bosons discussed in Sec.\ \ref{freemass}. 
The mass is treated as two-point interaction which leads by power-counting 
to a non-renormalizable theory in coordinate space. With standard 
differential renormalization it becomes necessary to adjust infinitely 
many constants. It will turn out that these constants coincide with 
the a priori fixed ratios of our approach. This makes it possible 
to recover the right result immediately within natural renormalization. 
This does not happen accidentally as can be shown in a general theorem.

\vspace{1ex}
\noindent{}The main application of this paper will be the $
\varphi ^{4}$-theory in Sec.\ \ref{fi4}. We focus our attention 
to the calculation of the two-point Green's function. It will turn 
out that the $\varphi ^{4}$-theory performs almost like made for 
our renormalization scheme: Most Feynman diagrams of a given order 
precisely match into a formula which allows to calculate their sum 
without evaluating single graphs. This enables us to calculate the 
two-point function up to five loops.

Finally the renormalization group is discussed. The $ \beta $-function 
and the anomalous dimension $\gamma $ are determined up to fourth 
and fifth order in the coupling, respectively.

\section{Definition of the renormalization scheme} 
\subsection{Comparison with differential renormalization\label
{diffreg}} 
We start with a generalization of the ideas presented in the introduction. 
Repeated application of the equation 
\begin{equation}
\label{3}\Box f\left( x^{2}\right) =\frac{4}{x^{2}{}} \frac{
\partial }{\partial x^{2}{}} x^{4}{} \frac{\partial }{\partial x
^{2}{}} f\left( x^{2}\right) 
\end{equation} 
leads to 
\begin{equation}
\label{4}\Box ^{n+1}{} \frac{1}{x^{2}{}} \ln\left( \frac{x^{2}}{
\Lambda ^{2}}\right) =-4^{n+1}n!\left( n+1\right) !\frac{1}{x^{
2n+4}{}} \hspace*{2ex}\hbox{ for }x\neq 0
\hspace{.6ex},\hspace{2ex}n=0,1,{\ldots}\hspace{.6ex}.
\end{equation} 
Note that these equations hold strictly only for $ x\neq 0$ and may 
be modified by $ \delta (x)$-terms (cf.\ Eq.\ (\ref{19})). In a 
renormalizable field theory one needs only a finite number of these 
equations ($ n=0,1$ for the $\varphi ^{4}$-theory), however it will 
turn out to be useful to look at the general case.

The function $ x^{-2n-4}$ has no well defined Fourier transform whereas 
$ x^{-2}\ln(x^{2}{} /\Lambda ^{2})$ has (cf.\ Eq.\ (\ref{I})). The 
differentially renormalized Fourier transform of the right hand 
side of Eq.\ (\ref{4}) is now determined by the left hand side with 
the Laplacian translated as $ -p^{2}$ \cite{Johnson}, 
\begin{equation}
\label{5}\int \limits _{{\rm diff.ren.\hspace{.38ex}}}\frac{d^{4
}x}{4\pi ^{2}{}} \frac{e^{ix\cdot p}}{x^{2n+4}{}} \equiv \frac{p
^{2n}}{\left( -4\right) ^{n+1}n!\left( n+1\right) !} \ln\left( 
\frac{p^{2}}{\bar{\Lambda }_{n}^{2}}\right) \hspace{.6ex}.
\end{equation} 
We have introduced different renormalization scales $ \bar
{\Lambda }_{n}$ for each $n$ to stress that they are integration 
constants which a priori are independent from each other and may 
differ by arbitrary positive factors. The $\Lambda _{n}$ are interpreted 
as renormalization scales.

Our analysis starts with the introduction of polar coordinates. 
\begin{equation}
\label{6}\int \frac{d^{4}x}{4\pi ^{2}{}} \frac{e^{ix\cdot p}}{x^{
2n+4}{}} =\frac{1}{\pi } \int _{0}^{\pi }d\vartheta \sin^{2}
\vartheta \int _{0}^{\infty }drr^{-2n-1}e^{ir|p|\cos\vartheta }
\hspace{.6ex},
\end{equation} 
where we have chosen the $z$-axis to be parallel to $p$. We evaluate 
the convergent $\vartheta $-integral and split the $r$-integral 
into a convergent part $ \int _{1}^{\infty }$ which is evaluated 
and a part $ \int _{0}^{1}$ that diverges at zero and has to remain 
unchanged.

The calculations are in principle straightforward but tedious \cite{Schnetz}. 
The result is ($ \sum _{0}^{-1}\equiv \sum _{1}^{0}\equiv 0$) 
\begin{eqnarray}
\label{888}\int \frac{d^{4}x}{4\pi ^{2}{}} \frac{e^{ix\cdot p}}{x
^{2n+4}{}} &=&\sum _{k=0}^{n-1}\frac{p^{2k}}{2\left( -4\right) ^{
k}k!\left( k+1\right) !} \left( \int _{0}^{1}\frac{dr}{r^{2n-2k
+1}{}} +\frac{1}{2n-2k} \right) \\ 
&&+\frac{2p^{2n}}{\left( -4\right) ^{n+1}n!\left( n+1\right) !} 
\left( -\int _{0}^{1}\frac{dr}{r} +\ln\left( \frac{e^C|p|}{2}
\right) -\frac{1}{2\left( n+1\right) } -\sum _{k=1}^{n}{} 
\frac{1}{k}\right) \hspace{.6ex}.\nonumber 
\end{eqnarray} 
From a mathematical point of view we want $p$-independent integrals 
to give $p$-independent results. So we are forced to make the following 
definitions in order to regain the result obtained by differential 
renormalization (\ref{5}), 
\begin{eqnarray}
\label{a}\int _{0}^{1}\frac{dr}{r^{2n-2k+1}{}} &=&-\frac{1}{2n-2
k}\hspace*{1cm}\hbox{\hspace{.38ex}and\hspace{.38ex}}\\ 
\label{b}\int _{0}^{1}\frac{dr}{r} &=&-\ln\Lambda _{n}-\frac{1}
{2n+2}-\sum _{k=1}^{n}{} \frac{1}{k}\hspace{.6ex}.
\end{eqnarray} 
Eq.\ (\ref{a}) is the analytic continuation of the formula 
\begin{equation}
\label{10}\int _{0}^{1}drr^{n}{} =\frac{1}{n+1}
\end{equation} 
to $ n<-1$. Eq.\ (\ref{b}) shows that within our approach we can 
not equate the renormalization scales $ \Lambda _{n}$ among each 
other. We find instead 
\begin{equation}
\label{c}\ln\Lambda _{n}=\ln\Lambda -\frac{1}{2n+2}-\sum _{k=1}^{
n}{} \frac{1}{k}
\end{equation} 
for some scale $\Lambda $. The difference $ \ln\Lambda _{i}-\ln
\Lambda _{j}$ for any $ i\neq j$ is a well-defined non-zero rational 
number. If one violates Eq.\ (\ref{c}) one changes the definition 
of convergent integrals or generates $p$-dependences from $p$-independent 
divergent integrals. In the differentially renormalized $
\varphi ^{4}$-theory $\Lambda _{0}$ and $\Lambda _{1}$ are usually 
equated which however does not destroy the self-consistency of the 
theory since it is incorporated in the freedom of choosing the renormalization 
scheme.

\vspace{1ex}
\noindent{}An overall factor in the renormalization constants is 
irrelevant, so we choose a renormalization scale $\Lambda $ according 
to 
\begin{equation}
\label{11}\int _{0}^{1}\frac{dr}{r} =-\ln\Lambda \hspace{.6ex}.
\end{equation}

Notice that the left hand side of this equation has no explicit $
\Lambda $ dependence. One assumes $ r^{-1}$ to have the implicit 
local $\Lambda $-term $ -\ln\Lambda \cdot \delta (r)$ in a similar 
way as the differentially renormalized version of $ x^{-4}$ acquires 
the local renormalization dependence $ -\frac{1}{4}\ln\Lambda _{
0}^{2}\cdot 4\pi ^{2}\delta ^{(4)}(x)$ (cf.\ Eq.\ (\ref{2})).

\subsection{First results\label{firstres}} 
'Natural renormalization' corresponds to differential renormalization 
with the $\Lambda _{n}$ defined via Eq.\ (\ref{c}). It gives a generalization 
of the usual definition of integrals.

The renormalization scale $\Lambda $ is kept for 'dimensional reasons'. 
If we integrate over dimensionful parameters then $\Lambda $ combines 
with other ln-terms to provide a scalar argument of the logarithms. 
$\Lambda $ is not a cutoff (notice that the integrals over higher 
order poles (\ref{a}) are $\Lambda $-independent), it is neither 
large nor small (cf.\ Sec.\ \ref{freemass}).

We summarize the above discussion by giving our definition for the 
singular Fourier transform ($ \bar{\Lambda }=2/e^C\Lambda $) 
\begin{equation}
\label{18}\int \frac{d^{4}x}{4\pi ^{2}{}} \frac{e^{ip\cdot x}}{x^{
2n+4}{}} =\frac{p^{2n}}{\left( -4\right) ^{n+1}n!\left( n+1
\right) !} \left( \ln\left( \frac{p^{2}}{\bar{\Lambda }^{2}}
\right) -\frac{1}{n+1}-2\sum _{k=1}^{n}{} \frac{1}{k} \right) 
\hspace{.6ex}.
\end{equation} 
Moreover we can derive this equation in the spirit of differential 
renormalization by Fourier transforming and translating the Laplacian 
$ \Box $ as $ -p^{2}$. But then we have to add $ \delta (x)$-terms 
in Eqs.\ (\ref{1}) and (\ref{4}) which are now uniquely fixed as 
\begin{equation}
\label{19}\Box ^{n+1}{} \frac{1}{x^{2}{}} \ln\left( \frac{x^{2}}
{\Lambda ^{2}}\right)  =-4^{n+1}n!\left( n+1\right) !\frac{1}{x^{
2n+4}{}} +\left( 2\sum _{k=1}^{n}{} \frac{1}{k}+\frac{1}{n+1} 
\right) \Box ^{n}\delta \left( x\right) \hspace{.6ex}.
\end{equation} 
The fundamental divergent integrals (\ref{10}) and (\ref{11}) are 
easily generalized to\footnote{In fact it is not possible to introduce 
different renormalization scales $\Lambda _{m}$ in Eq.\ (\ref{13}) 
as can e.g.\ be seen by comparing the $ (m+1)$-fold one-dimensional 
convolution of $ |r|^{-1}$ with the $ (m+1)$st power of the Fourier 
transform of $ |r|^{-1}$.} 
\begin{eqnarray}
\label{12}\int _{0}^{1}dr r^{n}\ln^{m}\left( r\right) &=&\frac{
\left( -1\right) ^{m}m!}{\left( n+1\right) ^{m+1}{}} 
\hspace{.6ex},\hspace{2ex}n\neq -1\hspace{.6ex},\hspace{2ex}m
\in {\ErgoBbb N}_{0}\hspace{.6ex},\\ 
\label{13}\int _{0}^{1}dr \frac{\ln^{m}\left( r\right) }{r} &=&-
\frac{\ln^{m+1}\left( \Lambda \right) }{m+1} 
\hspace{.6ex},\hspace{2ex}m\in {\ErgoBbb N}_{0}\hspace{.6ex}.
\end{eqnarray} 
So far we have only discussed singularities located at zero. By translation 
we can shift the poles to any point of ${\ErgoBbb R}$. At infinity 
however one could introduce a new renormalization scale $ 
\Lambda _{\infty }$ according to 
\begin{equation}
\label{20}\int _{1}^{\infty }\frac{dr}{r} =\ln\Lambda _{\infty }
\hspace{.6ex}.
\end{equation} 
$\Lambda $ should be proportional to $ \Lambda _{\infty }$ for dimensional 
reasons and it is very convenient\footnote{By Fourier transforms, 
e.g., singularities at zero are mapped to singularities at infinity. 
Eq.\ (\ref{18}) could also be obtained by an ($ n+2$)-fold convolution 
of $ p^{-2}$ (the Fourier transform of $ x^{-2}$). In this case 
the integrals are divergent at infinity and our result would depend 
on $\Lambda _{\infty }$. Comparison with (\ref{18}) leads to (\ref{D}) 
\cite{Schnetz}.} to set 
\begin{equation}
\label{D}\Lambda =\Lambda _{\infty }\hspace{.6ex}.
\end{equation} 
This allows us to generalize Eq.\ (\ref{20}) to 
\begin{eqnarray}
\label{14}\int _{1}^{\infty }drr^{n}\ln^{m}\left( r\right) &=&-
\frac{\left( -1\right) ^{m}m!}{\left( n+1\right) ^{m+1}{}} 
\hspace{.6ex},\hspace{2ex}n\neq -1\hspace{.6ex},\hspace{2ex}m
\in {\ErgoBbb N}_{0}\hspace{.6ex},\\ 
\label{15}\int _{1}^{\infty }dr\frac{\ln^{m}\left( r\right) }{r} 
&=&\frac{\ln^{m+1}\left( \Lambda \right) }{m+1} 
\hspace{.6ex},\hspace{2ex}m\in {\ErgoBbb N}_{0}\hspace{.6ex},
\end{eqnarray} 
and together with Eqs.\ (\ref{12}) and (\ref{13}) we get 
\begin{equation}
\label{16}\int _{0}^{\infty }drr^{n}\ln^{m}\left( r\right) =0
\hspace{.6ex}.
\end{equation} 
All the integrals defined so far can be summarized by the convention 
\begin{equation}
\label{01}0^{n}\equiv \infty ^{n}\equiv 0\hspace*{1ex}\forall n
\neq 0\hspace{.6ex},\hspace{2ex}\ln0\equiv \ln\infty \equiv \ln
\Lambda \hspace{.6ex}.
\end{equation} 
Note that these equations are symmetric under the interchange of 
zero and infinity which comes from the close connection to analytic 
continuation.

\vspace{1ex}
\noindent{}We close this section with some remarks on changing variables. 
Integrals that converge at infinity may be shifted by definition. 
However a naive rescaling $ r\mapsto ar$ in Eq.\ (\ref{11}) leads 
to 
\begin{equation}
\label{24}\int _{0}^{1/a}\frac{dar}{ar} =\int _{0}^{1}\frac{dr}{r
} +\int _{1}^{1/a}\frac{dr}{r} =-\ln\left( a\Lambda \right) 
\neq -\ln\Lambda \hspace{.6ex}.
\end{equation} 
To keep Eq.\ (\ref{11}) invariant under rescalings one has to treat 
the lower limit zero like a variable and write $ \int _{0}^{1}dr
/r=\int _{0/a}^{1/a}dr/r=\ln(1/a)-\ln(0/a)=-\ln\Lambda $. Or, equivalently, 
one rescales the renormalization scale $\Lambda $ according to $ 
\Lambda \mapsto \Lambda /a$. If, like in Eq.\ (\ref{10}), the integral 
does not depend on $\Lambda $, rescalings do not affect the result. 
For more complicated variable substitutions it is always appropriate 
to return to the original variables before one approaches the limits 
(cf.\ the bipyramide graph in Sec.\ \ref{bipyramide}).

\subsection{Relation to the theory of generalized functions\label
{genfunct}} 
We recognized already in the last section that Eq.\ (\ref{a}) can 
be understood in the context of analytic continuation. In order 
to include Eq.\ (\ref{11}) into this concept one has to 'care for 
dimensions' and multiply the integrand by the dimensionless factor 
$ (r/\Lambda )^\alpha $, $ \alpha \mapsto 0$, 
\begin{displaymath}
\int _{0}^{1}drr^{n}\left( \frac{r}{\Lambda }\right) ^\alpha =
\frac{\Lambda ^{-\alpha }}{n+\alpha +1} \hspace*{1ex}
\hspace{.6ex},
\end{displaymath} 
which gives $ (n+1)^{-1}$ for $ \alpha =0$, $ n\neq -1$ and $ 
\alpha ^{-1}-\ln\Lambda $ for $ n=-1$. If, according to Eq.\ (\ref{01}), 
we replace $\alpha ^{-1}$ by zero we are back at (\ref{11}). Note 
that analytic continuation is only correct if one uses the factor 
$ (r/\Lambda )^\alpha $ and if there exists an $\alpha $-region 
in ${\ErgoBbb C}$ where the integral converges. This prescription 
differs from dimensional regularization by the absence of the surface 
area $ \Omega _{\alpha +1}$. In general the $\alpha $-dependence 
of $ \Omega _{\alpha +1}$ cannot be compensated by a redefinition 
of the renormalization scale.

There are other contexts in which we can understand the renormalization 
scheme. Such are contour integrals in the complex plane or lattice 
theory which generalizes the Riemann sum prescription and eventually 
provides a purely numerical definition of divergent integrals \cite{Schnetz}. 
Here we present the relation to the theory of generalized functions.

Assume we are interested in an integral which contains the generalized 
function $f$ that is given as derivative of another generalized 
function $ F'=f$. With a test function $\varphi $ we obtain (e.g.\ 
\cite{Gelfand}) 
\begin{equation}
\label{23}\int _{-\infty }^{+\infty }dxf\left( x\right) \varphi 
\left( x\right) \equiv \left( f,\varphi \right) \equiv \left( F',
\varphi \right) \equiv -\left( F,\varphi '\right) \equiv -\int _{
-\infty }^{+\infty }dxF\left( x\right) \varphi '\left( x
\right) \hspace{.6ex}.
\end{equation} 
If $\varphi $ is sufficiently constant at the poles of $F$ the right 
hand side converges and can be used to define the integral on the 
left hand side.

Let us e.g.\ take $ f(x)=x^\lambda \Theta (x)$; $ \Theta (x)=1$ for 
$ x>0$ and $ \Theta (x)=0$ for $ x<0$. We choose $ \varphi =
\Theta (1-|x|)$ where the edges at $ x=\pm 1$ may be smoothed to 
be $ C^{\infty }$. In the limit where this becomes irrelevant we 
have for $ \lambda \neq -1$ 
\begin{equation}
\label{26a}\left( f,\varphi \right) =\int _{0}^{1}dxx^\lambda 
\equiv -\int _{0}^{\infty }dx\frac{x^{\lambda +1}}{\lambda +1} 
\left( -\delta \left( x-1\right) \right) =\frac{1}{\lambda +1}
\end{equation} 
which coincides with our renormalization rule (\ref{10}). 
For $ \lambda =-1$ we may take $ F(x)=\ln(x)\Theta (x)$ and get 
\begin{equation}
\label{27}\int _{0}^{1}\frac{dx}{x} =-\int _{0}^{\infty }dx\ln
\left( x\right) \left( -\delta \left( x-1\right) \right) =0
\end{equation} 
which is Eq.\ (\ref{11}) for $ \Lambda =1$. The same holds for Eqs.\ 
(\ref{12}) and (\ref{13}).

To see what happened with the renormalization scale $\Lambda $ we 
have to notice that the above calculation is ambiguous. There exist 
several functions $F$ which have the same derivative. On the real 
line they differ by a constant which is irrelevant since the test 
function $\varphi $ vanishes at $ \pm \infty $.

In general however the number of undetermined parameters equals the 
number of disconnected pieces of the integration domain. A singularity 
of the integrand $f$ at the origin splits ${\ErgoBbb R}$ into two 
disconnected parts $ {\ErgoBbb R}^-$ and $ {\ErgoBbb R}^+$. Each 
of the functions $ F(x)+C+D\Theta (x)$ is with the same right an 
integral of $ f(x)$ on the real line. However they give different 
results to the integrals (\ref{26a}) and (\ref{27}), 
\begin{eqnarray}
\label{28}&&\hspace*{-0.9cm}\int \limits _{0}^{1}\!\!dxx^
\lambda \equiv -\!\!\!\int \limits _{-\infty }^{\infty }\hspace
{-1ex} dx\left( C_\lambda +\Theta \left( x\right) \!\left( 
\frac{x^{\lambda +1}}{\lambda +1} +D_\lambda \right) \!\right) 
\left( \delta \left( x+1\right) -\delta \left( x-1\right) 
\right) =\frac{1}{\lambda +1} +D_\lambda ,\\ 
\label{29}&&\hspace*{-0.9cm}\int \limits _{0}^{1}\frac{dx}{x} 
\equiv -\int \limits _{-\infty }^{\infty }\hspace{-1ex} dx
\left( C_{-1}+\Theta \left( x\right) \left( \ln x+D_{-1}
\right) \right) \left( \delta \left( x+1\right) -\delta \left( x
-1\right) \right) =D_{-1}\hspace{.6ex},
\end{eqnarray} 
where all the $ C_\lambda $ and $ D_\lambda $ can be chosen separately.

The way out of this ambiguity is to change the topology of the integration 
domain. We can compactify ${\ErgoBbb R}$ to $ \overline {
{\ErgoBbb R}}$ by adding $\infty $ ($ =-\infty $). Since for $ 
\lambda <-1$ the integral over $ x^\lambda $ is well-defined at 
infinity we can define $F$ over $ \overline {{\ErgoBbb R}}
\backslash \{0\}$ which is again a connected domain with one integration 
constant. The $ D_\lambda \Theta (x)$ term is discontinuous at infinity 
and thus no longer allowed in Eq.\ (\ref{28}). The integral $ 
\int _{0}^{1}dxx^\lambda $ acquires again the unique value $ (
\lambda +1)^{-1}$.

For $ \lambda \ge -1$ the integral diverges at infinity and the gluing 
is not possible. However for $ \lambda >-1$ the integral is finite 
at $ x=0$ and can be defined on the connected domain $
{\ErgoBbb R}$. Just for the case $ \lambda =-1$ the ambiguity remains 
since the integral is divergent both at zero and infinity. The function 
$ \ln |x|$ can only be defined on $ {\ErgoBbb R}^-\cup 
{\ErgoBbb R}^+$ and one should keep the arbitrary constant $ D_{
-1}$ in Eq.\ (\ref{29}). With the more intuitive relabeling $ D_{
-1}=-\ln\Lambda $ we are back at Eq.\ (\ref{11}).

In practice the introduction of the renormalization scale $
\Lambda $ is a matter of convenience. It will turn out to be useful 
to have this parameter at hand. In principle one could set $ 
\Lambda =1$ using the standard theory of generalized functions and 
recover $\Lambda $ in physical results by getting the dimensions 
right in logarithmic terms.

\section{Applications} 
Now we turn to physical applications. In the following we are mainly 
concerned with four-dimensional integrals which we normalize according 
to 
\begin{equation}
\label{54}\int dx\equiv \int _{{\ErgoBbb R}^{4}}\frac{d^{4}x}{
\left( 2\pi \right) ^{2}{}} \hspace*{1cm}\hbox{\hspace{.38ex}and\hspace{.38ex}}
\hspace*{1cm}\delta \left( x\right) =\left( 2\pi \right) ^{2}
\delta ^{\left( 4\right) }\left( \vec{x}\right) \hspace{.6ex}.
\end{equation} 
This eliminates all irrelevant factors of $\pi $ from the theory. 
We get e.g.\ $ \int dx\delta (x)=1$ and $ \int dxe^{ip\cdot x}=
\delta (p)$. Analogously $n$-dimensional integrals are normalized 
by $ (2\pi )^{(-n/2)}$. The metric is always Euclidean.

\subsection{A toy example: the free massive bosonic theory\label
{freemass}} 
As a first test let us calculate the four-dimensional free massive 
boson propagator in coordinate space. The result is well known, 
\begin{eqnarray}
\Delta \left( x\right) &=&\int dp\frac{e^{-ip\cdot x}}{p^{2}+m^{
2}{}} =\frac{m}{|x|} K_{1}\left( m|x|\right) \nonumber \\ 
\label{51}&=&\frac{1}{x^{2}{}} +\frac{m^{2}}{4}\sum _{k=0}^{
\infty }\frac{\left( m^{2}x^{2}{} /4\right) ^{k}}{k!\left( k+1
\right) !} \left( \ln\left( \frac{m^{2}x^{2}}{4}\right) -\Psi 
\left( k+2\right) -\Psi \left( k+1\right) \right) 
\hspace{.6ex}.
\end{eqnarray} 
The propagator $\Delta $ is perfectly well-defined. However it is 
not analytic at $ m=0$ since the series contains logarithmic terms 
in $m$.

Now let us treat the mass $ (-m^{2})$ as a two-point interaction 
and study perturbation theory around $ m=0$. 
\begin{equation}
\bullet {\Etcompose{\Etcompose{\lline}{\raisebox{ .5pt}{$\lline
$}}}{\raisebox{-.5pt}{$\lline$}}}\bullet =\bullet \lline
\bullet +\Etcompose{\bullet \lline\bullet \lline\bullet }{
\raisebox{-1.3ex}{$\scriptstyle -m^{2}$}}\hspace{ 6.8ex} +
\Etcompose{\bullet \lline\bullet \lline\bullet \lline\bullet }{
\raisebox{-1.3ex}{$\scriptstyle -m^{2}\hspace{ 4ex} -m^{2}$}}
\hspace{ 6.8ex} +{\ldots}
\end{equation} 
The free propagator in four dimensions is 
\begin{equation}
\int dp\frac{1}{p^{2}{}} e^{-ip\cdot x}=\frac{1}{x^{2}{}} 
\hspace{.6ex},
\end{equation} 
and therefore 
\begin{equation}
\Delta \left( x\right) =\frac{1}{x^{2}{}} -m^{2}\int dx_{1}
\frac{1}{\left( x\!-\!x_{1}\right) ^{2}{}} \frac{1}{x_{1}^{2}{}
} +\left( -m^{2}\right) ^{2}\int \int dx_{1}dx_{2}\frac{1}{
\left( x\!-\!x_{1}\right) ^{2}{}} \frac{1}{\left( x_{1}\!-\!x_{
2}\right) ^{2}{}} \frac{1}{x_{2}^{2}{}} +{\ldots}\hspace{.6ex}.
\end{equation} 
Since the 'coupling' has mass-dimension the terms become more divergent 
with every order and the expansion is non-renormalizable. However 
we can treat the integrals according to our rules and obtain an 
unambiguous result which contains by construction only one renormalization 
scale $\Lambda $.

In order to evaluate the integrals we can use $ \Box (x-x_{1})^{
-2}=-\delta (x-x_{1})$ and Eq.\ (\ref{2}) to derive a recursive 
formula. Here it is even simpler to remember that the $n$-th term 
is the Fourier transform of $ p^{-2n}$. Eq.\ (\ref{18}) gives 
\begin{eqnarray}
\Delta _{{\rm nat.ren.\hspace{.38ex}}}\left( x\right) &=&\frac{1
}{x^{2}{}} +\sum _{k=0}^{\infty }\left( -m^{2}\right) ^{k+1}
\int dp\frac{1}{p^{2k+4}{}} e^{ip\cdot x}\nonumber \\ 
&=&\frac{1}{x^{2}{}} +\frac{m^{2}}{4}\sum _{k=0}^{\infty }
\frac{\left( m^{2}x^{2}{} /4\right) ^{k}}{k!\left( k+1\right) !
} \left( \ln\left( \frac{\Lambda ^{2}x^{2}}{4}\right) -\Psi 
\left( k+2\right) -\Psi \left( k+1\right) \right) \!.
\end{eqnarray} 
One could study the renormalization group by looking at rescalings 
of $\Lambda $. In fact comparison with Eq.\ (\ref{51}) shows that 
the situation is even simpler. We just have to equate $ 
\Lambda =m$ to obtain precisely the correct result. This does not 
happen accidentally as we will see in the next section.

\vspace{1ex}
\noindent{}The differentially renormalized result can be obtained 
by using Eq.\ (\ref{5}) instead of Eq.\ (\ref{18}), 
\begin{eqnarray}
\Delta _{{\rm diff.ren.\hspace{.38ex}}}\left( x\right) &=&\frac{
1}{x^{2}{}} +\frac{m^{2}}{4}\sum _{k=0}^{\infty }\frac{\left( m^{
2}x^{2}{} /4\right) ^{k}}{k!\left( k+1\right) !} \left( \ln
\left( \frac{\Lambda _{k}^{2}x^{2}}{4}\right) -2C\right) 
\hspace{.6ex}.
\end{eqnarray} 
It is necessary to adjust the infinitely many parameters $ 
\Lambda _{k}$ precisely according to the a priori settings (\ref{c}) 
of our scheme.

\vspace{1ex}
\noindent{}Dimensional regularization leads to a series in $ 4-n
$ with a simple pole, 
\begin{eqnarray}
\label{E}\Delta _{{\rm dim.reg.\hspace{.38ex}}}\left( x\right) &
=&\frac{1}{x^{2}{}} +\frac{m^{2}}{4}\sum _{k=0}^{\infty }\frac{
\left( m^{2}x^{2}{} /4\right) ^{k}}{k!\left( k+1\right) !} 
\left( \frac{2}{4-n}+\ln\left( \frac{\Lambda ^{2}x^{2}}{2}
\right) -\Psi \left( k+1\right) \right) \hspace{.6ex}.
\end{eqnarray} 
Since the series is non-renormalizable, it is not possible to renormalize 
by introducing a finite number of counter terms. If one nevertheless 
tries to follow a minimum subtraction prescription, one misses a 
term $ -\Psi (k+2)$ to obtain the correct result. In the next section 
we will present a general method that allows us to calculate this 
term.

\subsection{A theorem on singular expansions} 
Let us summarize what we did in the last section. We started from 
a well-defined integral $ \int dp\exp(-ip\cdot x)/(p^{2}+m^{2})
$ which we tried to expand into a series at $ m=0$. To this end 
we expanded the integrand into a power series $ \int dp\exp(-ip
\cdot x)\sum _{k=0}^{\infty }(-m^{2})^{k}p^{-2k-2}$. The interchange 
of the sum and the integral led us to the perturbation series in 
coordinate space $ \sum _{k=0}^{\infty }(-m^{2})^{k}\int dp\exp
(-ip\cdot x)p^{-2k-2}$. This interchange is obviously illegal. Firstly, 
the integrals diverge at $ p=0$. Secondly, we obtain a power series 
in $m$ and we know that the correct result has no such representation 
but contains logarithmic terms in $m$  (Eq.\ (\ref{51})). Although 
the integrand is analytic at $ m=0$ the integral is not. So necessarily 
the expansion is wrong and the diverging integrals reflect this 
fact. We want to study the issue how to reconstruct the true result 
from such an incorrect, singular expansion.

\vspace{1ex}
\noindent{}Let us slightly generalize the situation and look for 
the expansion of an integral $ I(a)=\int dxf(x,a)$ into a series 
at $ a=0$. The integrand has a Taylor (or Laurant) series $ f(x
,a)=\sum _{k}a^{k}f^{(k)}(x,0)/k!$, but in general we can not expect 
that the series of $ I(a)$ is given by the integrals over the coefficients 
$ f^{(k)}(x,0)$ since the integrals may diverge. We define 
\begin{equation}
\Delta I\left( a\right) =\int dxf\left( x,a\right) -\sum _{k}
\frac{a^{k}}{k!} \int dxf^{\left( k\right) }\left( x,0\right) 
\hspace{.6ex},
\end{equation} 
and conclude that $ \Delta I(a)$ will only be zero if $ I(a)$ is 
analytic at $ a=0$. So $ \Delta I(a)$ gives the part of the expansion 
of $ I(a)$ that can not be reached by standard perturbation theory.

We call $ \Delta I$ the non-perturbative part of the expansion. A 
priori we know almost nothing about it. However in many cases where 
$ I(a)$ is not analytic at $ a=0$ one can calculate $ \Delta I(a
)$ by the following theorem.

\pagebreak[3]

\noindent {\bf Theorem}. Assume the integrals $ \int dxf^{(k)}
(x,0)$ are regular at $ x\neq 0$. If there exists a neighborhood 
of $ x=0$, $ a=0$ where $f$ can be written as $ f=\sum _{\ell }f
_{\ell }$ with $ f_{\ell }(x,a)$ integrable at $ x=0$ and the $ f
_{\ell }^{(k)}(x,0)$ having the following properties 
\begin{eqnarray}
\label{41a}\lefteqn{f_{\ell }^{\left( k\right) }\left( |x|,0
\right) \propto |x|^{n\left( k,\ell \right) }\ln^{m\left( k,
\ell \right) }\left( |x|\right) \hspace*{2ex}\hbox
{\hspace{.38ex}with\hspace{.38ex}}}\\ 
\label{36a}&&\hbox{\hspace{.38ex}given }\ell :\hspace*{2ex}\lim_{
k\rightarrow \infty }n\left( k,\ell \right) /k <0\hbox{ or }f_{
\ell }^{\left( k\right) }\left( x,0\right) \equiv 0\hbox{ for almost 
all\footnotemark{} }k\hspace*{2ex}\hbox{\hspace{.38ex}and\hspace{.38ex}}
\\ 
\label{36b}&&\hbox{\hspace{.38ex}given }k:\hspace*{2ex}n\left( k
,\ell \right) \ge 0\hbox{ for almost all }\ell \hspace{.6ex},
\end{eqnarray} 
then 
\begin{equation}
\Delta I\left( a\right) =\sum _{\ell }\int dxf_{\ell }\left( x,a
\right) \hspace{.6ex}.
\end{equation}\footnotetext{all up to a finite number}

\vspace{1ex}
\noindent{}Note that all the integrals may diverge and have to be 
defined according to the rules given above. The range of integration 
can be $ {\ErgoBbb R}^{n}$ or $ {\ErgoBbb R}^+$, subsets can be 
taken into account by using step-functions.

\pagebreak[3]

\noindent {\bf Proof} (sketch). Without restriction we can assume 
that the support of $f$ is a little ball $ B_\varepsilon $ around 
$ x=0$ since the integrals over the remaining domain are regular 
and therefore do not contribute to $ \Delta I$. Moreover we can 
assume $ |a|$ to be small, so that we can write $f$ as a sum over 
$ f_{\ell }$. Since the $ f_{\ell }$ are integrable at $ x=0$ one 
gets for sufficiently small $\varepsilon $ 
\begin{eqnarray*}
\int _{B_\varepsilon }dxf\left( x,a\right) &=&\sum _{\ell }\int _
{B_\varepsilon }dxf_{\ell }\left( x,a\right) \\ 
&=&\sum _{\ell }\int _{{\ErgoBbb R}^{n}}dxf_{\ell }\left( x,a
\right) -\sum _{\ell }\int _{{\ErgoBbb R}^{n}\backslash B_
\varepsilon }dx\sum _{k}\frac{a^{k}}{k!} f_{\ell }^{\left( k
\right) }\left( x,0\right) \hspace{.6ex}.
\end{eqnarray*} 
In the second integral the singularity at $ x=0$ is excluded and 
(\ref{36a}) assures that the sum can be interchanged with the integral 
for small enough $a$, yielding 
\begin{displaymath}
\int _{B_\varepsilon }dxf\left( x,a\right) =\sum _{\ell }\int _{
{\ErgoBbb R}^{n}}dxf_{\ell }\left( x,a\right) +\sum _{\ell }
\sum _{k}\frac{a_{k}}{k!} \left( \int _{B_\varepsilon }-\int _{
{\ErgoBbb R}^{n}}\right) dxf_{\ell }^{\left( k\right) }\left( x
,0\right) \hspace{.6ex}.
\end{displaymath} 
Now we can use the central argument of the proof. The last integral 
over the entire $ {\ErgoBbb R}^{n}$ vanishes since (Eq.\ (\ref{41a})) 
it is proportional to $ \int _{0}^{\infty }dr\,r^N\ln^M(r)$ and 
all those integrals are zero in our renormalization scheme (Eq.\ 
(\ref{16})). We finally use Eq.\ (\ref{36b}) to interchange the 
second sum over $ \ell $ with the sum over $k$ and the integral. 
Therefore 
\begin{displaymath}
\Delta I\left( a\right) \equiv \int _{B_\varepsilon }dxf\left( x
,a\right) -\sum _{k}\frac{a^{k}}{k!} \int _{B_\varepsilon }dxf^
{\left( k\right) }\left( x,0\right) =\sum _{\ell }\int _{
{\ErgoBbb R}^{n}}dxf_{\ell }\left( x,a\right) \hspace{.6ex}.
\end{displaymath}\hfill $\Box $
\pagebreak[3] 

\vspace{1ex}
\noindent{}Now let us use the theorem to derive $ \Delta I(m^{2}
)$ of the scalar bosonic theory. We get $ f(p,m^{2})=e^{-ip
\cdot x}/(p^{2}+m^{2})$. Expanding the exponential yields $ f(p
,m^{2})=\sum _{\ell }(-ip\cdot x)^{\ell }{} /(\ell !(p^{2}+m^{2
}))\equiv \sum _{\ell }f_{\ell }(p,m^{2})$. Moreover $ f_{\ell 
}^{(k)}(p,0)=(-ip\cdot x)^{\ell }(-1)^{k}k!p^{-2-2k}{} /\ell !$ 
and $ f_{\ell }^{(k)}(|p|,0)\propto |p|^{\ell -2-2k}$ meets Eqs.\ 
(\ref{41a}), (\ref{36a}), (\ref{36b}). We can apply the theorem 
and obtain 
\begin{equation}
\label{F}\Delta I_{{\rm nat.ren.\hspace{.38ex}}}\left( m^{2}
\right) =\sum _{\ell =0}^{\infty }\int \frac{d^{4}x}{4\pi ^{2}
{}} \frac{\left( -ip\cdot x\right) ^{\ell }}{\ell !\left( p^{2}
+m^{2}\right) } =\sum _{\ell =0}^{\infty }\frac{\left( -i|x|
\right) ^{\ell }}{\ell !} \frac{1}{\pi } \int _{0}^{\pi }d
\vartheta \sin^{2}\vartheta \cos^{\ell }\vartheta \int _{0}^{
\infty }\frac{dp p^{\ell +3}}{p^{2}+m^{2}{}} .
\end{equation} 
The $\vartheta $-integral vanishes for odd $ \ell $. The divergent 
$p$-integral can be reduced to fundamental integrals as follows: 
$ \int _{0}^{\infty }dp p^{2\ell +3}{} /(p^{2}+m^{2})=\int _{0}^{
\infty }pdp ((p^{2}+m^{2})-m^{2})^{\ell +1}{} /(p^{2}+m^{2})$. With 
Eq.\ (\ref{16}) we obtain $ (-m^{2})^{\ell +1}\int _{0}^{
\infty }pdp/(p^{2}+m^{2}) =\frac{1}{2}(-m^{2})^{\ell +1}\ln(p^{
2}+m^{2})|^{\infty }_{0}=\frac{1}{2}(-m^{2})^{\ell +1}\ln(
\Lambda ^{2}{} /m^{2})$ (with Eq.\ (\ref{01})\footnote{More precisely 
$ \int _{0}^{\infty }pdp/(p^{2}+m^{2})=\int _{0}^{1}pdp/(p^{2}+m
^{2})+\int _{1}^{\infty }dp(p/(p^{2}+m^{2})-1/p)+\int _{1}^{
\infty }dp/p =\frac{1}{2}\ln(1+m^{-2})-\frac{1}{2}\ln(1+m^{2})+
\ln\Lambda =\ln\Lambda /m$.}). The result is proportional to $ 
\ln(\Lambda /m)$ and vanishes thus for $ \Lambda =m$. This confirms 
the explicit calculation of the last section.

The above theorem holds for any spacetime dimension. Hence it should 
as well be possible to apply it to the dimensionally regularized 
result and 'correct' Eq.\ (\ref{E}) by adding the non-perturbative 
part. We start with the $n$-dimensional analogon of Eq.\ (\ref{F}). 
All integrals are standard and one obtains 
\begin{eqnarray}
 \Delta I_{{\rm dim.reg.\hspace{.38ex}}}\left( m^{2}\right) &=&
\sum _{\ell =0}^{\infty }\frac{\left( -i|x|\right) ^{\ell }}{
\ell !} \frac{\Omega _{n-1}}{\left( 2\pi \right) ^{n/2}{}} 
\Lambda ^{4-n}{} \int _{0}^{\pi }d\vartheta \sin^{n-2}
\vartheta \cos^{\ell }\vartheta \int _{0}^{\infty }\frac{dp p^{
\ell +n-1}}{p^{2}+m^{2}{}} \nonumber \\ 
&=&\sum _{\ell =0}^{\infty }\frac{m^{2}\left( -x^{2}m^{2}
\right) ^{\ell }}{2^{\ell +2}\ell !} \left( \frac{m^{2}}{2
\Lambda ^{2}}\right) ^{\left( n-4\right) /2}\Gamma \left( 
\frac{4-n}{2}-\ell -1\right) \\ 
&=&\sum _{\ell =0}^{\infty }\frac{m^{2\ell +2}x^{2\ell }}{4^{
\ell +1}\ell !\left( \ell +1\right) !} \left( -\frac{2}{4-n}-
\Psi \left( \ell +2\right) +\ln\left( \frac{m^{2}}{2\Lambda ^{2
}}\right) \right) +{\cal O}\left( 4\hspace{-.5ex}-\hspace{-.5
ex} n\right) .\nonumber 
\end{eqnarray} 
Together with $ \Delta _{{\rm dim.reg.\hspace{.38ex}}}$ (Eq.\ (\ref{E})) 
the renormalization scale drops out and one obtains the full propagator.

\vspace{1ex}
\noindent{}So in the example of a free massive bosonic theory we 
do not have to go through the standard renormalization business. 
One can use the above theorem instead. The simplest way to expand 
the propagator is using natural renormalization, however dimensional 
regularization leads eventually to the same result.

In a realistic field theory with dimensionless coupling the situation 
is slightly different. The path integral is a priori ill-defined 
and the renormalization scale an intrinsic parameter of the theory 
(like e.g.\ in the integral $ \int _{0}^{1}\frac{dx}{x} =-\ln
\Lambda $). It makes no sense to equate the renormalization parameter 
$\Lambda $ with the coupling. However it is challenging to try to 
generalize the theorem to path integrals providing a non-perturbative 
but analytic definition of a quantum field theory.

\vspace{1ex}
\noindent{}Anyway the theorem on its own is useful in many elementary 
mathematical applications. Integrals like $ \int _{0}^bdx/(x^{n
}+a^{n})$, $ \int _{0}^{1}dx\ln^{m}(x/\Lambda )/(x+a)$, $ \int _{
0}^{\infty }dxe^{-bx}/(x+a)^{n}$, $ \int _{0}^{\infty }dxe^{-bx
-a/x}$, etc.\ can be expanded at $ a=0$ by virtue of the theorem 
\cite{Schnetz}.

\subsection{Fourier transforms\label{fourier}} 
Before we start to study $\varphi ^{4}$-theory it is useful to discuss 
Fourier transforms since many Feynman amplitudes are determined 
by multiplications and convolutions.

To this end we generalize the Fourier transforms discussed in the 
beginning (Eq.\ (\ref{18})). It is convenient to derive the result 
by analytic continuation. A straightforward calculation gives ($ 
\bar{\Lambda }=2/(e^C\Lambda )$) 
\begin{eqnarray}
\label{02}\int \!\frac{d^{4}x}{\left( 2\pi \right) ^{2}{}} 
\frac{e^{ip\cdot x}}{x^{2n+4}{}} \left( \frac{x^{2}}{\Lambda ^{
2}}\right) ^{\!-\alpha }\hspace{-2.5ex}&=&\hspace{-.5ex}\frac{p^{
2n}}{2^{2n+2}{}} \left( \frac{p^{2}}{\bar{\Lambda }^{2}}
\right) ^{\!\alpha }{} \frac{e^{-2C\alpha }\Gamma \left( -n-
\alpha \right) }{\Gamma \left( n+2+\alpha \right) } \equiv 
\frac{p^{2n}}{2^{2n+2}{}} \left( \frac{p^{2}}{\bar{\Lambda }^{2
}}\right) ^{\!\alpha }\sum _{\ell =-1}^{\infty }a_{n,\ell }
\alpha ^{\ell },\hspace*{ 3ex}\\ 
\hbox{\hspace{.38ex}where\hspace{.38ex}}\hspace*{3ex}\sum _{
\ell =-1}^{\infty }a_{n,\ell }\alpha ^{\ell }\hspace{-.5ex}&=&
\hspace{-.5ex}\frac{\Gamma \left( -n-\alpha \right) }{\Gamma 
\left( 1-\alpha \right) } \frac{\Gamma \left( 1+\alpha \right) 
}{\Gamma \left( n+2+\alpha \right) } \exp\left( \sum _{k=1}^{
\infty }\frac{2\zeta \left( 2k+1\right) }{2k+1}\alpha ^{2k+1}
\right) \hspace{.6ex}.
\end{eqnarray} 
Now it is easy to determine all Fourier transforms of the form $ x
^{-2n-4}\ln^{m}(x^{2}{} /\Lambda ^{2})$. To produce the logarithms 
we divide Eq.\ (\ref{02}) by $\alpha ^{m}$ and pick up the finite 
term in the $\alpha $-expansion: 
\begin{equation}
\int dx \frac{\ln^{m}\left( x^{2}{} /\Lambda ^{2}\right) }{x^{2n
+4}{}} e^{ip\cdot x}=\left( -1\right) ^{m}m!\frac{p^{2n}}{2^{2n
+2}{}} \sum _{\ell =-1}^{\infty }a_{n,\ell }g_{\ell -m}\left( 
\frac{p^{2}}{\bar{\Lambda }^{2}}\right) \hspace{.6ex},
\end{equation} 
where $ g_{\ell }(p^{2}{} /\bar{\Lambda }^{2})$ is the finite term 
of $ (p^{2}{} /\bar{\Lambda }^{2})^\alpha \alpha ^{\ell }$ at $ 
\alpha =0$. We obtain $ g_{\ell }=\ln^{|\ell |}(p^{2}{} /\bar
{\Lambda }^{2})/|\ell |!$ for $ \ell \le 0$. For $ \ell $ positive 
$ g_{\ell }(p^{2}{} /\bar{\Lambda }^{2})=0$ for all $ p\neq 0$. 
A more detailed calculation \cite{Schnetz} shows that $ p^{-4}g_{
1}(p^{2}{} /\bar{\Lambda }^{2})=\frac{1}{4}\delta (p)$.

Let us take e.g.\ $ n=1$ yielding $ \sum _{\ell =-1}^{\infty }a_{
1,\ell }\alpha ^{\ell }=\frac{1}{2}\alpha ^{-1}{} -\frac{5}{4}+
\frac{17}{8}\alpha +(-\frac{49}{64}+\frac{1}{3}\zeta (3))
\alpha ^{2}+{\cal O}(\alpha ^{3})$ and therefore (cf.\ Eq.\ (\ref{18})) 
\begin{eqnarray}
\label{G}\int dxe^{ip\cdot x}\frac{1}{x^{6}}&=&\frac{1}{16}p^{2}
\left( \frac{1}{2}\ln\left( \frac{p^{2}}{\bar{\Lambda }^{2}}
\right) -\frac{5}{4}\right) \hspace{.6ex},\\ 
\label{H}\int dxe^{ip\cdot x}\frac{\ln\left( x^{2}{} /\Lambda ^{
2}\right) }{x^{6}}&=&\frac{1}{16}p^{2}\left( -\frac{1}{4}\ln^{2
}\left( \frac{p^{2}}{\bar{\Lambda }^{2}}\right) +\frac{5}{4}\ln
\left( \frac{p^{2}}{\bar{\Lambda }^{2}}\right) -\frac{17}{8}
\right) \hspace{.6ex}.
\end{eqnarray} 
$ n=-1$ gives $ \sum _{\ell =-1}^{\infty }a_{-1,\ell }\alpha ^{
\ell }=1+\frac{2}{3}\zeta (3)\alpha ^{3}+{\cal O}(\alpha ^{5})$, 
thus 
\begin{equation}
\label{I}\int dxe^{ip\cdot x}\frac{1}{x^{2}{}} =\frac{1}{p^{2}{}
} \hspace{.6ex},\hspace{2ex}\hspace*{1cm}\int dxe^{ip\cdot x}
\frac{\ln\left( x^{2}{} /\Lambda ^{2}\right) }{x^{2}{}} =-
\frac{\ln\left( p^{2}{} /\bar{\Lambda }^{2}\right) }{p^{2}{}} 
\hspace{.6ex}.
\end{equation} 
With $ n=-2$ we finally obtain the standard formula $ \int dxe^{ip
\cdot x}=\delta (p)$. Less obvious is $ \int dxe^{ip\cdot x}\ln
(x^{2}{} /\Lambda ^{2})=-4p^{-4}+\delta (p)$.

\subsection{The massless $\varphi ^{4}$-theory\label{fi4}} 
The first serious test of the renormalization scheme is the discussion 
of the $\varphi ^{4}$-theory. Note that once the Feynman rules and 
the propagator are fixed the results are unique. There is no freedom 
to choose a certain subtraction scheme.

We keep our integral normalization of $ (2\pi )^{-2}$ which results 
in a rescaling of the coupling by $ (2\pi )^{2}$. So $g$ is related 
to the usual 'irrationalized' coupling via 
\begin{equation}
\frac{g}{4}=\frac{\lambda }{16\pi ^{2}{}} \hspace{.6ex}.
\end{equation}
\begin{figure}\centerline{\epsfbox{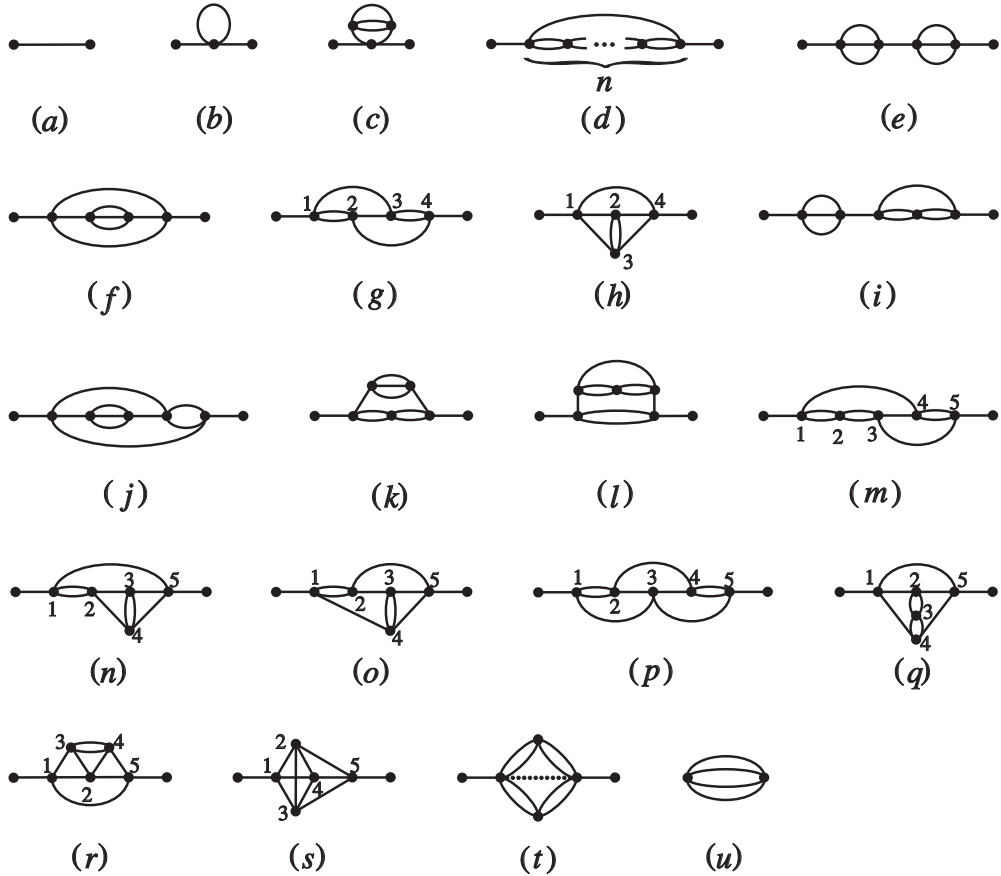}}
\caption{Feynman diagrams of the $\varphi ^{4}$-theory.}
\end{figure}
The Feynman diagrams we are concerned with are depicted 
in Fig.\ 1 ($ a$), {\dots}, ($ u$). The corresponding amplitudes 
are labeled by $ G_{a}$, {\dots}, $ G_u$. With natural renormalization 
we have the freedom to switch between coordinate and momentum space. 
However, most often it is convenient to start the calculation in 
coordinate space where, at least at higher loops, the Feynman rules 
are more transparent. The final result is given in momentum space 
to make it easier to compare it with other work.

\subsubsection{Simple results} 
The free propagator ($ a$) is given by $ G_{a}=p^{-2}$.

The loop in the diagram ($ b$) gives rise to a term $ \int dpp^{
-2}$ in momentum space or a term $ (x-x)^{-2}{} =0^{-2}$ in coordinate 
space. Both expressions are set to zero in our scheme: $ G_b=0$. 
In this aspect it behaves like dimensional regularization.

More generally, diagrams that contain tadpole insertions give zero 
and can be dropped. This remains true for any number of internal 
lines the tadpole may have: $ G_c=0$. The reason is that in a massless 
theory a tadpole insertion can only give rise to a number times 
a momentum conserving $\delta $-function. On the other hand it has 
dimension $ p^{2}$ and the only number with this scaling property 
is zero. In our renormalization scheme $ \bar{\Lambda }^{2}$ occurs 
only in combination with logarithms.

Moreover, due to translation invariance and Eq.\ (\ref{16}), all 
vacuum bubbles vanish: $ G_u=0$. So only connected diagrams contribute 
to the two-point function. Altogether this reduces the number of 
relevant Feynman diagrams considerably.

Diagram ($ d$) for $ n=1$ is the sunset diagram. It was already calculated 
in the last section. The triple line gives $ x^{-6}$ which transforms 
into momentum space as $ (\frac{p}{4})^{2}(\frac{1}{2}\ln(p/
\bar{\Lambda } )^{2}- \frac{5}{4})$. Together with the two external 
legs and the symmetry factor $\frac{1}{6}$ we obtain 
\begin{equation}
G_{d,1}=\left( \frac{g}{4}\right) ^{2}{} \frac{1}{p^{2}{}} 
\left( \frac{1}{12}\ln\left( \frac{p^{2}}{\bar{\Lambda }^{2}}
\right) -\frac{5}{24}\right) \hspace{.6ex}.
\end{equation}

\subsubsection{Chain graphs} 
We call diagrams of type ($ d$) chain graphs. To any order there 
exists one chain graph and, if we disregard the vanishing diagrams 
with tadpoles, the only remaining diagrams up to three loops are 
chain graphs.

It is possible to calculate chain graph amplitudes for any $n$ by 
Fourier transformation. In momentum space the series of bubbles 
gives rise to the $n$-th power of $ \frac{1}{4}(-\ln(p/\bar
{\Lambda })^{2}+1)$ (cf.\ Eq.\ (\ref{18})). A final convolution 
with $ p^{-2}$ (use Eq.\ (\ref{3}) as suggested in Sec.\ \ref{freemass}) 
provides the result as an $n$-th order polynomial in $ \ln(p/
\bar{\Lambda })^{2}$ with purely rational coefficients. Including 
the symmetry factors and the external legs we obtain for $ n
\ge 2$ (the case $ n=1$ has an extra symmetry which changes the 
symmetry factor from $ 2^{-n}$ to $\frac{1}{6}$) 
\begin{equation}
G_{d,n}=\left( \frac{g}{4}\right) ^{n+1}{} \frac{1}{p^{2}{}} 
\frac{n!}{2^{n+1}{}} \sum _{k=0}^{n}\left( -1\right) ^{k}\frac{
\ln^{n-k}\left( p/\bar{\Lambda }\right) ^{2}}{\left( n-k
\right) !} \sum _{\ell =0}^{k}\frac{2-2^{\ell -k}}{\ell !} 
\hspace{.6ex}.
\end{equation} 
There is a nice way to compile this result by a generating function. 
If we multiply $ G_{d,n}$ with the factor $ a_{n}=3^{n-2}$ for $ n
\ge 2$, $ a_{1}=1$, it reproduces the leading logarithms of the 
full $\varphi ^{4}$ two-point function correctly. The result may 
be seen as some approximation to the propagator. We get 
\begin{equation}
\label{03}\sum _{n=1}^{\infty }\frac{a_{n}G_{d,n}}{n!} =\frac{g}
{9p^{2}{}} \left( \frac{\left( p^{2}{} /e\bar{\Lambda }^{2}
\right) ^{3g/8}}{9\left( 1+g/4\right) ^{2}-1} -\frac{1}{8}
\right) \hspace{.6ex}.
\end{equation} 
We easily read off 
\begin{eqnarray}
\label{75}\hspace{-3ex} G_{d,2}&\hspace{-1.1ex}=\hspace{-1.1ex}&
\left( \frac{g}{4}\right) ^{3}\!\frac{1}{p^{2}{}} \!\left( 
\frac{1}{8}\ln^{2}\!\left( \frac{p^{2}}{\bar{\Lambda }^{2}}
\right) -\frac{5}{8}\ln\!\left( \frac{p^{2}}{\bar{\Lambda }^{2}
}\right)  +\frac{15}{16}\right) \\ 
\label{76}\hspace{-3ex} G_{d,3}&\hspace{-1.1ex}=\hspace{-1.1ex}&
\left( \frac{g}{4}\right) ^{4}\!\frac{1}{p^{2}{}} \!\left( 
\frac{1}{16}\ln^{3}\!\left( \frac{p^{2}}{\bar{\Lambda }^{2}}
\right)  -\frac{15}{32}\ln^{2}\!\left( \frac{p^{2}}{\bar
{\Lambda }^{2}}\right)  +\frac{45}{32}\ln\!\left( \frac{p^{2}}{
\bar{\Lambda }^{2}}\right)  -\frac{109}{64}\right) \\ 
\label{J}\hspace{-3ex} G_{d,4}&\hspace{-1.1ex}=\hspace{-1.1ex}&
\left( \frac{g}{4}\right) ^{5}\!\frac{1}{p^{2}{}} \!\left( 
\frac{1}{32}\ln^{4}\!\left( \frac{p^{2}}{\bar{\Lambda }^{2}}
\right)  -\frac{5}{16}\ln^{3}\!\left( \frac{p^{2}}{\bar
{\Lambda }^{2}}\right)  +\frac{45}{32}\ln^{2}\!\left( \frac{p^{
2}}{\bar{\Lambda }^{2}}\right)  -\frac{109}{32}\ln\!\left( 
\frac{p^{2}}{\bar{\Lambda }^{2}}\right) +\frac{239}{64}\right) 
\!.
\end{eqnarray}

\subsubsection{Four loops} 
Before we start with the analysis of four and five loops a word of 
caution is in order. In general it is not sufficient to define the 
integral over generalized functions for defining a field theory, 
since also products of generalized functions appear. In principle 
e.g.\ one has to consider terms like $ x^{2}\delta (x)$ since they 
might give finite contributions after multiplication with $ x^{
-2}$.

In the following we do not care about such terms. The main message 
of the next two subsections is to show that there is a miraculous 
matching of Feynman amplitudes in the natural renormalization scheme 
that makes calculations easy. This matching is not affected by the 
above problems nor are the leading logarithms of the results. This 
is confirmed by the existence of the renormalization group equation 
studied in Sec.\ \ref{rengroup}.

\vspace{1ex}
\noindent{}Diagrams ($ e$), ($ f$), ($ g$), and ($ h$) remain to 
be evaluated. $ G_e$ is basically the square of $ G_{d,1}$. 
\begin{equation}
G_e=\left( \frac{g}{4}\right) ^{4}{} \frac{1}{p^{2}{}} \left( 
\frac{1}{144}\ln^{2}\left( \frac{p}{\bar{\Lambda }}\right) ^{2}
{} -\frac{5}{144}\ln\left( \frac{p}{\bar{\Lambda }}\right) ^{2}
{} +\frac{25}{576}\right) \hspace{.6ex}.
\end{equation} 
$ G_f$ can be calculated with Fourier transforms. Adding propagators 
from the interior loop to the exterior lines we obtain (see Eqs.\ 
(\ref{G})--(\ref{I})) 
\begin{eqnarray*}
&&\hspace{-0.65cm} \frac{1}{x^{6}{}} \stackrel{{\cal F}}{
\longrightarrow } \frac{p^{2}}{4^{2}{}} \left( \frac{1}{2}\ln
\left( \frac{p}{\bar{\Lambda }}\right) ^{2}\!\!-\frac{5}{4}
\right)  \stackrel{p^{-4}}{\longrightarrow } \frac{1}{4^{2}p^{2
}{}} \left( \frac{1}{2}\ln\left( \frac{p}{\bar{\Lambda }}
\right) ^{2}\!\!-\frac{5}{4}\right)  \stackrel{{\cal F}}{
\longrightarrow } \frac{1}{4^{2}x^{2}{}} \left( -\frac{1}{2}\ln
\left( \frac{x}{\Lambda }\right) ^{2}\!\!-\frac{5}{4}\right) \\
&&\hspace{-0.65cm} \stackrel{x^{-4}}{\longrightarrow } \frac{1}{4
^{2}x^{6}{}} \left( -\frac{1}{2}\ln\left( \frac{x}{\Lambda }
\right) ^{2}\!\!-\frac{5}{4}\right)  \stackrel{{\cal F}}{
\longrightarrow } \frac{p^{2}}{4^{4}{}} \left( \frac{1}{8}\ln^{
2}\left( \frac{p}{\bar{\Lambda }}\right) ^{2}\!\!-\frac{5}{8}\ln
\left( \frac{p}{\bar{\Lambda }}\right) ^{2}\!\!+\frac{17}{16}{} 
-\frac{5}{8}\ln\left( \frac{p}{\bar{\Lambda }}\right) ^{2}\!\!+
\frac{25}{16}\right) .
\end{eqnarray*} 
Together with the external lines and the symmetry factor $ 
\frac{1}{12}$ one gets 
\begin{equation}
G_f=\left( \frac{g}{4}\right) ^{4}{} \frac{1}{p^{2}{}} \left( 
\frac{1}{96}\ln^{2}\left( \frac{p^{2}}{\bar{\Lambda }^{2}}
\right)  -\frac{5}{48}\ln\left( \frac{p^{2}}{\bar{\Lambda }^{2}
}\right)  +\frac{7}{32}\right) \hspace{.6ex}.
\end{equation} 
We are left with two diagrams each of which cannot be calculated 
by Fourier transformation. Both have the same symmetry factor $\frac{1}{4}$. 
This makes it possible to use a formula which is specific to four 
dimensions.

\vspace{1ex}
\centerline{\epsfbox{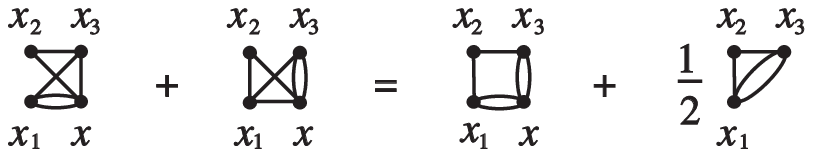}}\vspace{-1.8cm}
\begin{equation}
\label{100a}
\end{equation}\vspace{.6cm}

\noindent{}This equation holds up to a total derivative proportional 
to 
\begin{equation}
\frac{\partial }{\partial x^{\mu }{}} \left( \frac{1}{\left( x-x_{
1}\right) ^{2}{}} \frac{x^{\mu }-x_{2}^{\mu }}{\left( x-x_{2}
\right) ^{2}{}} \frac{1}{\left( x-x_{3}\right) ^{2}{}} -\frac{x^{
\mu }-x_{1}^{\mu }}{\left( x-x_{1}\right) ^{4}{}} \frac{1}{
\left( x-x_{3}\right) ^{2}}\right) 
\end{equation} 
 (notice that $ \frac{\partial }{\partial x^{\mu }{}} \frac{x^{
\mu }-x_{1}^{\mu }}{(x-x_{1})^{4}{}} =\frac{1}{2}\delta (x-x_{1
})$). After integration over $x$ the total derivative vanishes and 
we obtain:

\vspace{3ex}
\epsfbox{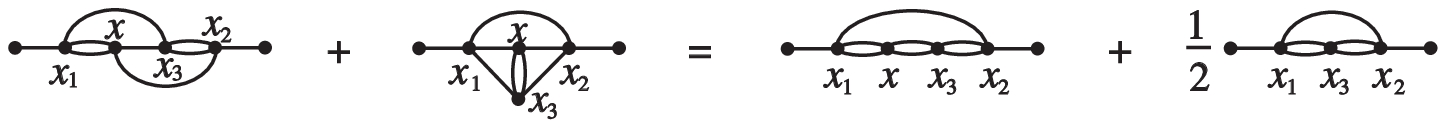}\vspace{-1.6cm}
\begin{equation}
\label{100}
\end{equation}\vspace{.2cm}

\noindent{}Thus we need not solve each of the complicated diagrams 
($ g$) and ($ h$) separately\footnote{We do this in the next subsection. 
The single results will be more complicated than the sum. Each of 
the amplitudes ($ g$) and ($ h$) has a $ \zeta (3)$-dependence that 
cancels in the sum.}. Their sum is equal to two chain diagrams.

Eq.\ (\ref{100a}) can be interpreted as integration by parts which 
also proved to be useful within dimensional regularization \cite{Chetyrkin}. 
However only in natural renormalization it allows one to calculate 
the sum of diagrams without evaluating single graphs. One should 
take this as a hint that calculating single diagrams is in general 
not an appropriate method to evaluate higher order perturbation 
theory. All diagrams (or at least groups of diagrams) of a given 
order should be treated as a unit and calculated together. This 
strategy will be even more useful in the next section. 
\begin{equation}
G_g+G_h=\left( \frac{g}{4}\right) ^{4}{} \frac{1}{p^{2}{}} 
\left( \frac{1}{8}\ln^{3}\left( \frac{p^{2}}{\bar{\Lambda }^{2}
}\right)  -\frac{19}{16}\ln^{2}\left( \frac{p^{2}}{\bar
{\Lambda }^{2}}\right)  +\frac{65}{16}\ln\left( \frac{p^{2}}{
\bar{\Lambda }^{2}}\right)  -\frac{169}{32}\right) 
\hspace{.6ex}.
\end{equation}

\subsubsection{Five loops.\label{fiveloops}} 
Apart from the trivial diagram 
\begin{equation}
G_{i}=\left( \frac{g}{4}\right) ^{5}{} \frac{1}{p^{2}{}} \left( 
\frac{1}{96}\ln^{3}\left( \frac{p^{2}}{\bar{\Lambda }^{2}}
\right)  -\frac{5}{64}\ln^{2}\left( \frac{p^{2}}{\bar{\Lambda }^{
2}}\right)  +\frac{5}{24}\ln\left( \frac{p^{2}}{\bar{\Lambda }^{
2}}\right) -\frac{25}{128}\right) 
\end{equation} 
and the five loop chain graph, Eq.\ (\ref{J}), ten diagrams have 
to be evaluated. These graphs split into three classes: (1) Diagrams 
that can be solved by Fourier transforms ($j$ -- $l$). Let us call 
such diagrams Fourier graphs. (2) Diagrams that can be reduced to 
Fourier graphs via integration by parts ($m$ -- $r$), and (3) the 
nonplanar diagram ($s$) that we call the bipyramide graph. 
\paragraph{Fourier graphs.} 
Every graph that reduces under the replacement of multiple lines 
($ \bullet \hspace{-1.3ex}=\hspace{-.5ex}=\hspace{-1.3ex}
\bullet $, $ \bullet \hspace{-1.3ex}\equiv \hspace{-.5ex}
\equiv \hspace{-1.3ex}\bullet $) and iterated lines ($ \bullet 
\hspace{-1ex}-\hspace{-1ex}-\hspace{-1ex}\bullet \hspace{-1ex}-
\hspace{-1ex}-\hspace{-1ex}\bullet $) by a simple line ($ 
\bullet \hspace{-1ex}-\hspace{-1ex}-\hspace{-1ex}\bullet $) to the 
free propagator can be solved with Fourier transforms. The calculations 
are analogous to the evaluation of diagram ($f$\/) in the last section. 
Including the respective symmetry factors $\frac{1}{12}$, $\frac{1}{24}$, 
$\frac{1}{8}$ we get 
\begin{eqnarray}
G_{j}&\!=\!&\left( \frac{g}{4}\right) ^{5}{} \frac{1}{p^{2}{}} 
\left( \frac{1}{96}\ln^{3}\!\left( \frac{p^{2}}{\bar{\Lambda }^{
2}}\right)  -\frac{25}{192}\ln^{2}\!\left( \frac{p^{2}}{\bar
{\Lambda }^{2}}\right)  +\frac{97}{192}\ln\!\left( \frac{p^{2}}
{\bar{\Lambda }^{2}}\right) -\frac{269}{384}\right) 
\hspace{.6ex},\\ 
\label{75a}G_{k}&\!=\!&\left( \frac{g}{4}\right) ^{5}{} \frac{1}
{p^{2}{}} \left( \frac{1}{144}\ln^{3}\!\left( \frac{p^{2}}{\bar
{\Lambda }^{2}}\right)  -\frac{5}{64}\ln^{2}\!\left( \frac{p^{2
}}{\bar{\Lambda }^{2}}\right)  +\frac{59}{192}\ln\!\left( 
\frac{p^{2}}{\bar{\Lambda }^{2}}\right) -\frac{173}{384}+\frac{
1}{36}\zeta \left( 3\right) \right) \!,\\ 
G_l&\!=\!&\left( \frac{g}{4}\right) ^{5}{} \frac{1}{p^{2}{}} 
\left( \frac{1}{96}\ln^{3}\!\left( \frac{p^{2}}{\bar{\Lambda }^{
2}}\right)  -\frac{5}{32}\ln^{2}\!\left( \frac{p^{2}}{\bar
{\Lambda }^{2}}\right)  +\frac{57}{64}\ln\!\left( \frac{p^{2}}{
\bar{\Lambda }^{2}}\right) -\frac{209}{128}+\frac{1}{24}\zeta 
\left( 3\right) \right) \hspace{.6ex}.
\end{eqnarray} 
For future use we calculate the improper four loop $\varphi ^{4}
$-diagram ($t$) where the dotted line means a '$ (-1)$-fold' propagator 
$ (x-y)^{+2}$. The result is (with a symmetry factor of $\frac{1}{32}$) 
\begin{equation}
\label{75b}G_t=\left( \frac{g}{4}\right) ^{4}{} \frac{1}{p^{2}{}
} \left( \frac{1}{48}\ln^{3}\left( \frac{p^{2}}{\bar{\Lambda }^{
2}}\right)  -\frac{5}{32}\ln^{2}\left( \frac{p^{2}}{\bar
{\Lambda }^{2}}\right)  +\frac{17}{32}\ln\left( \frac{p^{2}}{
\bar{\Lambda }^{2}}\right) -\frac{49}{64}+\frac{1}{12}\zeta 
\left( 3\right) \right) \hspace{.6ex}.
\end{equation} 
\paragraph{Integration by parts.} 
We determine the following symmetry factors: ($m$): $\frac{1}{8}$, 
($n$): $\frac{1}{8}$, ($o$): $\frac{1}{4}$, ($p$): $\frac{1}{4}$, 
($q$): $\frac{1}{8}$, ($r$): $\frac{1}{2}$.

The idea is to use Eq.\ (\ref{100a}) to relate the above graphs among 
each other. Sometimes it will be necessary to multiply Eq.\ (\ref{100a}) 
by $ (x_{1}-x_{2})^{2}$, $ (x_{1}-x_{3})^{2}$, or $ (x_{2}-x_{3
})^{2}$. Since these factors are independent of $x$ they do not 
affect partial integration with respect to $x$. However, if these 
factors do not combine with propagators $ (x_{1}-x_{2})^{-2}$, etc., 
one obtains improper $\varphi ^{4}$-graphs like diagram ($t$). In 
most cases it is possible to eliminate those graphs by a second 
application of Eq.\ (\ref{100a}). In the following table we denote 
first the graph we start from, then the variables which correspond 
to $ (x,x_{1},x_{2},x_{3})$ in Eq.\ (\ref{100a}) (according to Fig.\ 
1), occasionally the variables of a second application of Eq.\ (\ref{100a}), 
and finally the resulting equation including the symmetry factors. 
\begin{equation}
\hbox{\hspace{.38ex}
\begin{tabular}{cccl} 
graph&$ \left( x,x_{1},x_{2},x_{3}\right) $&$ \left( x,x_{1},x_{
2},x_{3}\right) $&equation\\\hline  
($p$)&(2,1,4,3)&---&$ G_p+G_o=2G_{m}-\frac{1}{2}gG_g$\\ 
($r$)&(3,4,1,2)&---&$ \frac{1}{2}G_r=G_q-\frac{1}{4}gG_h$\\ 
($r$)&(3,4,2,1)&(2,4,1,5)&$ \frac{1}{2}G_r+2G_{n}+2G_{m}=4G_{d,4
}-\frac{1}{2}gG_h-gG_{d,3}$\\ 
($m$)&(2,3,5,1)&(4,3,1,5)&$ 2G_q+G_o+2G_{m}=4G_{d,4}-\frac{1}{2}gG_h
-gG_{d,3}$\\\hline  
($g$)&(2,1,4,3)&---&$ G_h=-G_g+2G_{d,3}-\frac{1}{2}gG_{d,2}$\hspace*{1cm}(Eq.\ 
(\ref{100}))\\ 
($g$)&(2,1,3,4)&---&$ G_g=4G_t-\frac{1}{4}gG_{d,2}$\\ 
---&---&---&$ gG_t=12G_{k}+\frac{5}{32}G_{d,2}$\hspace*{2ex}(Eq.\ 
(\ref{75}), (\ref{75a}), (\ref{75b}))
\end{tabular}\hspace{.38ex}}
\end{equation} 
The last equation can explicitly be checked by looking at the amplitudes. 
We recognize that there are only four equations to evaluate six 
five loop diagrams. However summing up the first four equations 
gives 
\begin{equation}
2G_{m}+2G_{n}+2G_o+G_p+G_q+G_r=8G_{d,4}-\left( g/4\right) \cdot 
\left( 2G_g+5G_h+8G_{d,3}\right) \hspace{.6ex}.
\end{equation} 
With the last three equations in the table we can express the left 
hand side completely in terms of Fourier amplitudes of the $
\varphi ^{4}$-theory 
\begin{eqnarray}
\lefteqn{2G_{m}+2G_{n}+2G_o+G_p+G_q+G_r\hspace*{1ex}=
\hspace*{1ex}8G_{d,4}+36G_{k}-18\left( g/4\right) G_{d,3}+
\frac{29}{2}\left( g/4\right) ^{2}G_{d,2}}\nonumber \\ 
&&\hspace{-4ex}=\left( \frac{g}{4}\right) ^{5}{} \frac{1}{p^{2}
{}} \left( \frac{1}{4}\ln^{4}\!\left( \frac{p^{2}}{\bar
{\Lambda }^{2}}\right) -\frac{27}{8}\ln^{3}\!\left( \frac{p^{2}
}{\bar{\Lambda }^{2}}\right) +\frac{299}{16}\ln^{2}\!\left( 
\frac{p^{2}}{\bar{\Lambda }^{2}}\right) -\frac{809}{16}\ln\!
\left( \frac{p^{2}}{\bar{\Lambda }^{2}}\right) +\frac{1853}{32}
+\zeta \left( 3\right) \right) .\nonumber \\ 
&&
\end{eqnarray} 
The graphs ($m$), ($n$), ($o$) are not symmetric under interchange 
of the external legs. Therefore we have to count them twice in the 
two-point function and the left hand side becomes exactly the combination 
we want to calculate. 
\paragraph{The bipyramide graph.\label{bipyramide}} 
The bipyramide graph ($s$) is the first non-planar two-point graph 
and commonly regarded as the most complicated five-loop diagram. 
It was first calculated in 1981 within dimensional regularization 
by K.G. Chetyrkin and F.V. Tkachov \cite{Chetyrkin}. Recently it 
was analyzed within differential renormalization by V.A. Smirnov 
\cite{Smirnov3}.

So it is a good candidate to test the power of our calculation scheme. 
We work in coordinate space. It is convenient to introduce a quaternionic 
notation. The inversion of a quaternion $x$ is given by $ x
\mapsto \frac{1}{x}$ which can be understood in the four vector 
language as inversion of the length of $x$ ($ |x| \mapsto |x|^{
-1}$) and a reflection at the $z$-axis (the direction of the unit 
quaternion 1). The square $ (x-y)^{2}$ becomes the square of the 
absolute $ |x-y|^{2}$, however we stick to the brackets in the following 
calculation to keep the notation more transparent.

The variables $ (x,a,b,c,y)$ correspond to $ (1,2,3,4,5)$ in Fig.\ 
1.\ We have to calculate the following integral 
\begin{equation}
\label{78}\int dadbdc\frac{1}{\left( x-a\right) ^{2}\left( x-b
\right) ^{2}\left( x-c\right) ^{2}{}} \frac{1}{\left( a-b
\right) ^{2}\left( a-c\right) ^{2}\left( b-c\right) ^{2}{}} 
\frac{1}{\left( a-y\right) ^{2}\left( b-y\right) ^{2}\left( c-y
\right) ^{2}{}} \hspace{.6ex}.
\end{equation} 
The external legs are amputated, they can easily be added in the 
end.

The integral is convergent at infinity (it is logarithmically divergent 
at $ a=b=c=y$ and $ a=b=c=z$) and therefore the integration variables 
$ a,b,c$ can be shifted by $y$. With $ z=x-y$ we have 
\begin{equation}
I\left( z\right) =\int dadbdc \frac{1}{a^{2}b^{2}c^{2}{}} \frac{
1}{\left( a-b\right) ^{2}\left( a-c\right) ^{2}\left( b-c
\right) ^{2}{}} \frac{1}{\left( z-a\right) ^{2}\left( z-b
\right) ^{2}\left( z-c\right) ^{2}{}} \hspace{.6ex}.
\end{equation} 
With the inversions $ a'=\frac{1}{a}$, $ b'=\frac{1}{b}$, $ c'=
\frac{1}{c}$ one obtains ($ d^{4}a=a'^{-8}d^{4}a'$) 
\begin{eqnarray*}
&&\int da'db'dc' \frac{1}{a'^{6}b'^{6}c'^{6}{}} \frac{1}{\left( z
-\frac{1}{a'}\right) ^{2}\left( z-\frac{1}{b'}\right) ^{2}
\left( z-\frac{1}{c'}\right) ^{2}{}} \frac{1}{\left( \frac{1}{a'} 
-\frac{1}{b'}\right) ^{2}\left( \frac{1}{a'} -\frac{1}{c'}
\right) ^{2}\left( \frac{1}{b'} -\frac{1}{c'}\right) ^{2}{}}\\ 
&&\frac{1}{z^{6}{}} \int da'db'dc' \frac{1}{\left( a'-\frac{1}{
z}\right) ^{2}\left( b'-\frac{1}{z}\right) ^{2}\left( c'-\frac{
1}{z}\right) ^{2}{}} \frac{1}{\left( b'-a'\right) ^{2}\left( c'-a'
\right) ^{2}\left( c'-b'\right) ^{2}{}} \hspace{.6ex}.
\end{eqnarray*} 
A shift $ a''=a'-\frac{1}{z}$, $ b''=b'-\frac{1}{z}$, $ c''=c'-
\frac{1}{z}$ yields 
\begin{equation}
\label{79b}\frac{1}{z^{6}{}} \int da''db''dc'' \frac{1}{a''^{2}b''
^{2}c''^{2}{}} \frac{1}{\left( a''-b''\right) ^{2}\left( a''-c''
\right) ^{2}\left( b''-c''\right) ^{2}{}} \hspace{.6ex}.
\end{equation} 

\vspace{1ex}
\noindent{}It seems that we have lost the $z$-dependence in the integral. 
However, since the integral is still divergent, this is not the 
case as we will see soon.

We finally use the rescaling $ b''=a''u$, $ c''=a''v$ to obtain 
\begin{equation}
\label{79}\frac{1}{z^{6}{}} \int \frac{da''}{a''^{4}{}} \int du
\int dv\frac{1}{\left( 1-u\right) ^{2}u^{2}\left( u-v\right) ^{
2}v^{2}\left( v-1\right) ^{2}{}} \hspace{.6ex}.
\end{equation} 
The $u$- and $v$-integral is finite and gives a positive number. 
It can be evaluated using Gegenbauer polynomial techniques (e.g.\ 
\cite{Chetyrkin2}, \cite{Johnson}) with the result\footnote{Most 
efficiently one uses the identity $ |uv|/(u-v)^{2}{} =\frac{1}{
\pi } \int _{-\infty }^{\infty }dP (|u|/|v|)^{iP}\sum _{n=1}^{
\infty }nC_{n-1}(\hat{u}\cdot \hat{v})\cdot (n^{2}+P^{2})^{-1}$ 
and the orthogonality relation $ \frac{1}{2\pi ^{2}{}} \int d
\hat{x}nC_{n-1}(\hat{y}\cdot \hat{x})mC_{m-1}(\hat{x}\cdot \hat
{z})=\delta _{n,m}nC_{n-1}(\hat{y}\cdot \hat{z})$ to obtain $ 
\frac{1}{\pi } \hspace{-1ex}\int _{-\infty }^{\infty }dP\sum _{
n=1}^{\infty }n^{2}(n^{2}+P^{2})^{-3}$.} $ \frac{3}{8}\zeta (3)
$. 
The angular integral in $ \int da'' /a''^{4}$ gives $ 2\pi ^{2}$. 
Including the normalization $ \frac{1}{4\pi ^{2}}$ (Eq.\ (\ref{54})) 
one is left with the radial integral $ \frac{1}{2}\int _{0}^{
\infty }d|a''|/|a''|=\frac{1}{2}\ln(|a''|)|_{a''=0}^{a''=
\infty }$. If we would have started with an integral in the $ a''$-variable 
this integral would give zero. However as discussed at the end of 
Sec.\ \ref{firstres} it is now essential to reintroduce the original 
variable $a$ before approaching the limits. Since $ a''=\frac{1
}{a}-\frac{1}{z}$, 
\begin{equation}
\int \limits _{0}^{\infty }\frac{d|a''|}{2|a''|} =\left. \frac{1
}{2}\ln\!\left( \left| \frac{1}{a}-\frac{1}{z}\right| \right) 
\right| _{a=z}^{a=0}\!\!=\left. \frac{1}{2}\ln\!\left( \frac{|z
-a|}{|az|}\right) \right| _{a=z}^{a=0}\!\!=\frac{1}{2}\left( \ln
\!\left( \frac{1}{\Lambda }\right)  -\ln\!\left( \frac{\Lambda 
}{z^{2}}\right)  
\right) =\frac{1}{2}\ln\!\left( \frac{z^{2}}{\Lambda ^{2}}
\right) .
\end{equation} 
Collecting all pieces we have finally found 
\begin{equation}
I\left( z\right) =\frac{3}{16}\zeta \left( 3\right) \frac{1}{z^{
6}{}} \ln\left( \frac{z^{2}}{\Lambda ^{2}}\right)  
\hspace{.6ex}.
\end{equation} 
The transformation to momentum space is given by Eq.\ (\ref{H}). 
Including the external legs and the symmetry factor $\frac{1}{6}$ 
we obtain 
\begin{equation}
G_{m}=\left( \frac{g}{4}\right) ^{5}{} \frac{1}{p^{2}{}} \left( 
\frac{1}{2}\zeta \left( 3\right) \ln^{2}\left( \frac{p^{2}}{
\bar{\Lambda }^{2}}\right) - \frac{5}{2}\zeta \left( 3\right) \ln
\left( \frac{p^{2}}{\bar{\Lambda }^{2}}\right)  +\frac{17}{4}
\zeta \left( 3\right) \right) \hspace{.6ex}.
\end{equation} 
Comparing with dimensional regularization \cite{Chetyrkin} gives 
the minimum coincidence that both results are proportional to $
\zeta $(3). It is not possible to be more precise since in \cite{Chetyrkin} 
only the singular part was calculated. Note that the techniques 
we used can not be generalized to dimensions different from four.

It is also hard to compare our result with the one gained by differential 
renormalization in \cite{Smirnov3} since the author restricted himself 
to regularize the amplitude and did not evaluate the rather complicated 
integrals over the internal variables.

\subsubsection{The two-point Green's function} 
Collecting all results from the last sections we obtain for the full 
propagator of the $\varphi ^{4}$-theory 
\begin{eqnarray}
G\left( p\right) &=&\frac{1}{p^{2}{}} \left( 1+\left( \frac{g}{4
}\right) ^{2}\left( \frac{1}{12}\ln\left( \frac{p^{2}}{\bar
{\Lambda }^{2}}\right) -\frac{5}{24}\right) +\left( \frac{g}{4}
\right) ^{3}\left( \frac{1}{8}\ln^{2}\left( \frac{p^{2}}{\bar
{\Lambda }^{2}}\right) -\frac{5}{8}\ln\left( \frac{p^{2}}{\bar
{\Lambda }^{2}}\right) +\frac{15}{16}\right) \right. \nonumber 
\\ 
&&+\left( \frac{g}{4}\right) ^{4}\left( \frac{3}{16}\ln^{3}
\left( \frac{p^{2}}{\bar{\Lambda }^{2}}\right) -\frac{59}{36}\ln
^{2}\left( \frac{p^{2}}{\bar{\Lambda }^{2}}\right)  +\frac{1535
}{288}\ln\left( \frac{p^{2}}{\bar{\Lambda }^{2}}\right) -\frac{
121}{18}\right) \nonumber \\ 
&&+\left( \frac{g}{4}\right) ^{5}\left( \frac{9}{32}\ln^{4}
\left( \frac{p^{2}}{\bar{\Lambda }^{2}}\right) -\frac{1045}{288
}\ln^{3}\left( \frac{p^{2}}{\bar{\Lambda }^{2}}\right) +\left( 
\frac{3733}{192}+\frac{1}{2}\zeta \left( 3\right) \right) \ln^{
2}\left( \frac{p^{2}}{\bar{\Lambda }^{2}}\right) \right. 
\nonumber \\ 
&&\label{994}-\left. \left( \frac{1643}{32}+\frac{5}{2}\zeta 
\left( 3\right) \right) \ln\left( \frac{p^{2}}{\bar{\Lambda }^{
2}}\right) +\frac{3697}{64}+\frac{383}{72}\zeta \left( 3
\right) \right) +{\cal O}\left( g^{6}\right) \hspace{.6ex}.
\end{eqnarray} 
Comparison with differential renormalization \cite{Johnson} 
\begin{equation}
\label{995}G_{{\rm diff.ren.\hspace{.38ex}}}\left( p\right) =
\frac{1}{p^{2}{}} \left( 1+\left( \frac{g}{4}\right) ^{2}{} 
\frac{1}{12}\ln\left( \frac{p}{\bar{\Lambda }}\right) ^{2}\!+
\left( \frac{g}{4}\right) ^{3}\left( \frac{1}{8}\ln^{2}\left( 
\frac{p}{\bar{\Lambda }}\right) ^{2}\!-\frac{3}{8}\ln\left( 
\frac{p}{\bar{\Lambda }}\right) ^{2}\right) +{\ldots}\right) 
\end{equation} 
shows that only the leading logarithms coincide. Notice that one 
never gets $ \ln$-independent terms in \cite{Johnson}.

If $ \bar{\Lambda }$ is rescaled according to $ \ln\bar
{\Lambda } \mapsto \ln\bar{\Lambda }+1$ in Eq.\ (\ref{995}) the 
logarithmic terms coincide with that of Eq.\ (\ref{994}). The $ 
\ln$-independent terms can be adjusted via a momentum independent 
rescaling by $ 1-\frac{3}{24}(g/4)^{2}{} +\frac{11}{16}(g/4)^{3
}$. However, at this point it is not clear whether the differences 
disappear after appropriate redefinitions also at higher orders.

\subsubsection{The renormalization group\label{rengroup}} 
It is possible to extract the $ \beta $-function and the anomalous 
dimension $\gamma $ from the two-point function alone if one assumes 
that $ \beta $ and $\gamma $ are independent of $ \bar
{\Lambda }$. Moreover the existence of a renormalization group equation 
is a non-trivial test for the renormalization scheme. Comparing 
the coefficients in 
\begin{equation}
\left( \frac{\partial }{\partial \ln\bar{\Lambda }} +\beta 
\left( \frac{g}{4}\right) \frac{\partial }{\partial \left( g/4
\right) } +2\gamma \left( \frac{g}{4}\right) \right) G\left( 
\frac{g}{4},\bar{\Lambda },p\right) =0
\end{equation} 
yields 
\begin{eqnarray}
\beta \left( \frac{g}{4}\right) &=&3\left( \frac{g}{4}\right) ^{
2}{} -\frac{17}{3}\left( \frac{g}{4}\right) ^{3}+\left( \frac{7
9}{4}+12\zeta \left( 3\right) \right) \left( \frac{g}{4}
\right) ^{4}+{\cal O}\left( g^{5}\right) \hspace{.6ex},\\ 
\gamma \left( \frac{g}{4}\right) &=&\frac{1}{12}\left( \frac{g}
{4}\right) ^{2}{} -\frac{5}{96}\left( \frac{g}{4}\right) ^{4}{} 
+\frac{191}{192}\left( \frac{g}{4}\right) ^{5}+{\cal O}\left( g^{
6}\right)  \hspace{.6ex}.
\end{eqnarray} 
The first two terms of $ \beta $ and the first term of $\gamma $ 
are standard. The coefficient in front of the $ \zeta (3)$-term 
also coincides with other schemes \cite{Johnson}, \cite{Smirnov}. 
However e.g.\ the vanishing third order and the $\zeta $(3)-independent 
fifth order term of $\gamma $ is specific to our scheme. In differential 
renormalization \cite{Johnson} one obtains $ \beta (g/4)=3(g/4)^{
2}{} -\frac{17}{3}(g/4)^{3}+(31+12\zeta (3))(g/4)^{4}+{\ldots}$, 
$ \gamma (g/4)=\frac{1}{12}(g/4)^{2}{} -\frac{3}{8}(g/4)^{3}+
{\ldots}$.

\section{Results and outlook} 
A new renormalization scheme was proposed. It provides all amplitudes 
fully renormalized, it has no explicit cutoff or counterterms and 
allows to keep the spacetime dimension fixed. The scheme defines 
all integrals in an unambiguous way, it thus corresponds to a definite 
choice of a subtraction prescription.

The renormalization scheme emerges from differential renormalization 
by an a priori fixing of all integration constants at their mathematically 
most natural values. It is closely related to the theory of generalized 
functions.

We demonstrated how to use this renormalization scheme if applied 
to the toy problem of a two-point (mass) interaction in coordinate 
space. Although this theory is non-renormalizable by power-counting 
it was possible to recover the correct result within our scheme. 
A theorem was presented that allowed us in a more general framework 
to reconstruct the full result from such a singular expansion. With 
this theorem it was possible (but more complicated) to regain the 
true result even for the dimensionally regularized toy model which 
failed to give the correct perturbation series.

\vspace{1ex}
\noindent{}The main application of our scheme was the $\varphi ^{
4}$-theory. Equations that are very special to four dimensions and 
to our renormalization prescription enabled us to calculate the 
two-point Green's function up to five loops (Eq.\ (\ref{994})). 
Most remarkable was the observation that at (four) five loops the 
diagrams are organized in such a way that a (one-) two-fold underdetermined 
system of linear equations could be solved for the sum over certain 
diagrams. This made the evaluation of many single graphs needless.

We were left with the nonplanar five-loop graph which could as well 
be calculated analytically in our renormalization scheme. It is 
obvious that the dimension of spacetime plays a crucial role in 
the calculation of the bipyramide graph (as it does for the matching 
of diagrams via integration by parts). Only in four dimensions the 
coupling becomes dimensionless. The resulting conformal symmetry 
was used via the inversion $ a\mapsto a^{-1}$ as the most essential 
step in the evaluation of the integrals.

The two-point function was compatible with the renormalization group 
and it was possible to extract the $ \beta $-function up to fourth 
and the anomalous dimension $\gamma $ up to fifth order in the coupling.

\vspace{1ex}
\noindent{}

For future work the idea of grouping certain classes of diagrams 
and calculating their sum without referring to single graphs appears 
especially promising to us. We expect that the matching of amplitudes 
persists to some extent at higher orders. In this way perturbation 
theory could be simplified and even analytical results beyond the 
fifth order may be possible. (Recent calculations confirm this for 
the sixth order of $\varphi ^{4}$-theory.) Most desirable would 
be to find the general structure that organizes the amplitudes to 
groups that can be evaluated via integration by parts. General questions 
of renormalizability and the problems related to the multiplication 
of generalized functions have to be investigated more carefully.

A goal of obvious importance is the application of natural renormalization 
to gauge theories. In general one has to avoid conflicts between 
Ward identities (reflecting gauge symmetry) and the renormalization 
scheme. This problem is already present in two dimensions and can 
be solved by using the transverse (Landau) gauge. The Schwinger 
model can be solved within this framework by summing up the whole 
perturbation series (e.g.\ the fermion correlation function) \cite{Schnetz}. 
The key tool is, similar to the $\varphi ^{4}$-theory, to calculate 
whole classes of Feynman diagrams without evaluating single amplitudes. 
In four dimensions first results are promising, however for QED 
we have not yet found how to group diagrams to simplify the calculations.

\section*{Acknowledgement} 
I thank Prof.\ M. Thies and Prof.\ F. Lenz for the pleasant and fruitful 
cooperation. Further I am grateful to Prof.\ K. Johnson and Prof.\ 
A. Bassetto for interesting and helpful discussions.

\end{document}